\documentclass[pdfpagelabels=false]{aa}  

\usepackage{graphicx}
\usepackage{txfonts}
\usepackage{orcidlink}

\usepackage{hyperref}	
\hypersetup{colorlinks=true,linkcolor=blue,citecolor=blue,filecolor=blue,urlcolor=blue}

\usepackage{siunitx}
\begin{document} 

\title{Projection effects in star-forming regions}
\subtitle{I. Nearest-neighbour statistics and observational biases}

\authorrunning{Barnes et al.}
\author{A.~T.~Barnes \inst{\ref{eso}}\orcidlink{0000-0003-0410-4504} \and 
        K.~Morii \inst{\ref{cfa}}\orcidlink{0000-0002-6752-6061} \and 
        J.~E. Pineda \inst{\ref{mpe}}\orcidlink{0000-0002-3972-1978} \and 
        R. J.~Parker \inst{\ref{shef}}\orcidlink{0000-0002-1474-7848}\thanks{Royal Society Dorothy Hodgkin fellow} \and 
        E.~Schisano \inst{\ref{inaf}}\orcidlink{0000-0003-1560-3958} \and 
        A.~Traficante \inst{\ref{iaps-inaf}}\orcidlink{0000-0003-1665-6402} \and 
        E.~Redaelli \inst{\ref{eso}}\orcidlink{0000-0002-0528-8125} \and
        K.~Immer \inst{\ref{eso}} \and
        J.~D.~Henshaw \inst{\ref{mpia}} \orcidlink{0000-0001-9656-7682} \and
        P.~Sanhueza \inst{\ref{UTokyo}}\orcidlink{0000-0002-7125-7685} \and
        F.~Motte \inst{\ref{Grenoble}}\orcidlink{0000-0003-1649-8002} \and 
        A.~Hacar \inst{\ref{Vienna}}\orcidlink{0000-0001-5397-6961}
        }

\institute{
\label{eso} European Southern Observatory (ESO), Karl-Schwarzschild-Stra{\ss}e 2, 85748 Garching, Germany  \and
\label{cfa} Center for Astrophysics | Harvard \& Smithsonian, 60 Garden Street, Cambridge, MA 02138, USA \and 
\label{mpe} Max-Planck-Institut für extraterrestrische Physik, Giessenbachstrasse 1, D-85748 Garching, Germany \and 
\label{shef} Astrophysics Research Cluster, School of Mathematical and Physical Sciences, The University of Sheffield, Hounsfield Road, Sheffield S3 7RH, UK \and
\label{inaf} INAF-Istituto di Astrofisica e Planetologia Spaziale, Via Fosso del Cavaliere 100, I-00133 Roma, Italy \and
\label{iaps-inaf} IAPS-INAF, Via Fosso del Cavaliere, 100, I-00133 Rome, Italy \and
\label{mpia} Max Planck Institut fur Astronomie, Heidelberg, Germany \and
\label{UTokyo} Department of Astronomy, School of Science, The University of Tokyo, 7-3-1 Hongo, Bunkyo, Tokyo 113-0033, Japan \and
\label{Grenoble} Univ. Grenoble Alpes, CNRS, IPAG, 38000 Grenoble, France \and
\label{Vienna} Department of Astrophysics, University of Vienna, T¨urkenschanzstrasse 17, 1180, Vienna (Austria)
}

\date{Received 28/11/2025; accepted 19/01/2026}

\abstract{
Stars are formed as molecular clouds fragment into networks of dense cores, filaments,
and sub-clusters. The characteristic spacing of these dense cores is therefore a
key observable imprint of the underlying fragmentation physics and is often
compared to theoretical scales such as the Jeans or sonic length. Nearest–neighbour
(NN) statistics are widely used to measure this spacing, yet they are derived from
projected 2D positions, while fragmentation unfolds in three
dimensions. Using a hierarchy of spherical and fractal toy models, we show
that the standard geometric de-projection factor of $4/\pi\simeq1.27$ is inadequate
because two effects operate together: (1) Projection not only foreshortens
separations but also rewires the NN network, creating artificial 2D links
between sources that are widely separated in 3D. (2) Finite angular resolution
introduces beam blending, which merges close neighbours and inflates the apparent
separations. 
We quantify these opposing biases with Monte Carlo experiments spanning a wide
range of morphologies, sample sizes, and resolutions, parametrized by the number
of independent beams across the field of view. From this parameter space analysis we derived a simple
empirical correction factor that depends on both the number of identified objects
and the effective resolution. For small samples or coarsely resolved data
($N\!\lesssim\!10$ or $\lesssim$10 beams across the field), the intrinsic mean NN
spacing exceeds the projected value by only $\sim$20–40\%, while for well-sampled,
well-resolved maps ($N\!\gtrsim\!100$ and $\gtrsim$30–50 beams across the field),
the true 3D separations are typically larger than the observed 2D spacings by a
factor of approximately two. In practice, this calibration allows observers to take a
measured 2D NN spacing and estimate a corresponding 3D value by applying a
resolution- and sample-size–dependent multiplicative factor, with typical
morphology-driven systematic uncertainties on the order of 30–40\%. We compare this
framework to observed and simulated core populations and show how it modifies
inferences about preferred fragmentation scales. This work is a first step towards
quantifying projection bias in core separations. We deliberately omitted additional
complexities such as sensitivity limits, background confusion, and incomplete
field of view, and we outline paths forward via synthetic observations,
hydrodynamic simulations, and velocity-resolved datasets to build a more complete
framework for interpreting 2D spacing statistics in star-forming regions.
}

\keywords{}
\maketitle

\section{Introduction}
\label{sec_int}

Stars rarely form in isolation (e.g. \citealp{Lada2003}). 
They emerge from the hierarchical fragmentation
of molecular clouds into networks of dense filaments and compact cores that
collapse to produce binary systems, associations, and clusters (e.g. see recent reviews 
\citealp{Pineda2023, Hacar2023, Offner2023, Wright2023, Chevance2023}). 
The spatial distribution of these cores can encode the physics that regulates fragmentation—
revealing whether gravity, turbulence, magnetic fields, or stellar feedback
dominate in shaping the birth environment of stars. 
Measuring the characteristic core spacing therefore offers a probe of the
fragmentation scale, a strategy already exploited in early studies of regularly
spaced condensations along filaments and in nearby clouds
\citep[e.g.][]{Schneider1979, Hartmann2001} as well as spacings within young stellar objects (YSOs)
and more evolved stellar populations \citep[e.g.][]{Gutermuth2009, Kirk2011, Schmeja2011}. 
These spacings are often compared to theoretical benchmarks such as the Jeans length
\citep[e.g.][]{Jeans1902, Larson1985, Inutsuka1997} or the turbulent and
magnetically modified Jeans scales
\citep{Padoan2002, Federrath2012, Hennebelle2019}. 
Because the nascent stellar population inherits its clustering properties and
multiplicity from the core distribution, understanding these separations is
essential for linking cloud structure to the origin of stellar systems.
However, almost all observational measurements are made in projection, where
the 3D geometry of fragmentation is collapsed onto the plane of
the sky. Projection effects are therefore one of the key factors limiting our
ability to draw robust conclusions about how molecular clouds fragment to form
stars.

In nearby, predominantly low-mass star-forming regions, regular spacings between
dense condensations along filaments have long been reported, and they are often found to be
of order the thermal Jeans length of the parent gas structure
\citep[e.g.][]{Schneider1979,Hartmann2002,Schmalzl2010,Palau2015,Tafalla2015,
Teixeira2016,Kainulainen2017}. 
The advent of Atacama Large Millimeter/submillimeter Array (ALMA), which combines sub-arcsecond resolution with high sensitivity to
dust continuum and molecular-line tracers, has extended such studies to more
distant, high-mass environments. Recent surveys of infrared dark clouds (IRDCs)
and massive clumps likewise have found core separations comparable to the thermal
Jeans length rather than to the turbulent Jeans scale
\citep[e.g.][]{Beuther2015b,Beuther2018,Palau2018,Liu2019,Sanhueza2019,
Lu2020,Beuther2021,Ishihara2024}. 
These results suggest that thermal fragmentation dominates in many cold, dense
environments. However, in more evolved feedback-dominated regions, turbulence
may play a stronger role (see \citealp{Rebolledo2020, Jiao2023, Avison2023}; despite caveats in identifying and characterising cores \citealp{Pouteau2023}).
Other studies have proposed a hierarchical fragmentation sequence, from clump to
sub-clump to cores, based on double-peaked or multi-scale separation distributions
\citep{Teixeira2016,Kainulainen2016,Henshaw2016a,Henshaw2017,Palau2018,Pokhrel2018,Svoboda2019,
Rosen2020,Zhang2021,Thomasson2022,Thomasson2024}. 
Recent studies also suggest that the spatial distribution of dense cores evolves dynamically over time: Cores initially form at relatively large separations but progressively migrate and concentrate towards the central regions of the forming cluster \citep{Traficante2023}. This decrease in spacing during the early phases results in an overall increase of the observed level of fragmentation (e.g. \citealp{Schisano2025}).

\begin{figure}
    \centering
    \includegraphics[width=\linewidth]{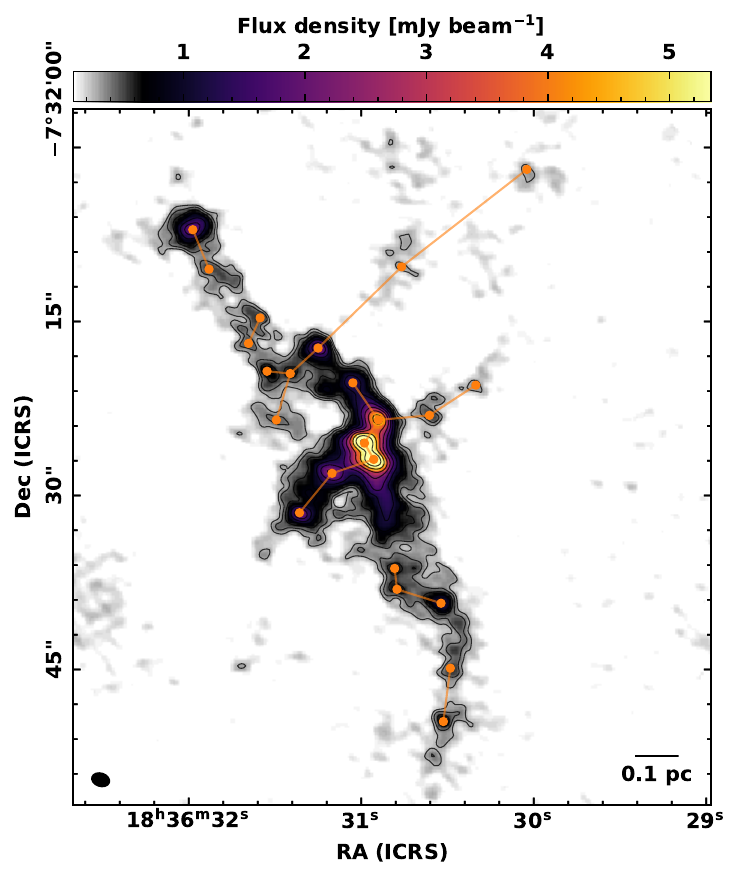}
    \caption{
    Example of a 1.3\,mm continuum image from the ASHES survey \citep{Morii2023}, overlaid with the NN graph (orange lines). 
    The background colour scale and contours show the ALMA 12\,m plus 7\,m continuum emission, with contour levels at 
    $3\times2^{\,n}\sigma$ ($n = 0, 1, 2, \dots$), where $\sigma = 9.5\times10^{-5}$\,Jy\,beam$^{-1}$ is the rms noise level. 
    Detected core positions are marked by orange circles, and the scale bar in the lower left corresponds to 0.1\,pc at the assumed distance of 5.5\,kpc.
    }
    \label{fig:ashes_cont}
\end{figure}

Despite this progress, most analyses are limited to case studies or in spatial coverage, and a statistical framework for interpreting core separations in three dimensions
remains lacking.  New large–area ALMA programs such as the
ALMA evolutionary study of high-mass protocluster formation in the GALaxy (ALMAGAL; \citealp{Molinari2025}), ALMA initial mass function (ALMA-IMF; \citealp{Motte2022}), ALMA Survey of 70\,$\mu$m dark High-mass clumps in Early Stages (ASHES; 
\citealp{Sanhueza2019, Morii2024}, and ALMA Infrared dark cloud (ALMA-IRDC; \citealp{Barnes2021c}) surveys now provide the necessary
statistical reach to test fragmentation theories across environments.  However,
the interpretation of their measured core separations relies on understanding
how projection onto the plane of the sky biases the underlying 3D structure.  Figure~\ref{fig:ashes_cont} illustrates this for one of the fields of ASHES: Dense cores (orange circles) identified in the continuum map are connected by their nearest neighbour (NN; orange lines).

A major complication arises because we can only observe the 2D
projection of a 3D distribution of dense cores.
Projection effects necessarily alter the apparent separations, and hence the
measured spacing distributions, in ways that can bias comparisons to intrinsic
physical scales. Without a reliable correction, there is a risk of systematically
under- or overestimating the physical processes inferred from these length scales.
Such projection effects have been discussed in several recent observational studies
\citep[e.g.][]{Henshaw2016a, Sanhueza2019, Traficante2023, Ishihara2024, Morii2024}, typically in the context of applying a simple geometric de-projection factor.

For example, under the assumption of isotropy, a core–core separation of true
three–dimensional length, $\ell_{\rm 3D}$, making an angle, $\theta$, to the
line of sight projects to (see \citealp{Ishihara2024})
\begin{equation}
  \ell_{\rm 2D} \;=\; \ell_{\rm 3D}\,\sin\theta.
\end{equation}
For an isotropic distribution of orientations, the polar angle, $\theta$, is
distributed according to $p(\theta)\,d\theta = \tfrac{1}{2}\sin\theta\,d\theta$
on $\theta\in[0,\pi]$. The expected value of $\sin\theta$ is then
\begin{equation}
  \big\langle \sin\theta \big\rangle
  \;=\; \frac{1}{2}\int_{0}^{\pi}\sin^{2}\theta\,d\theta
  \;=\; \frac{\pi}{4},
\end{equation}
yielding
\begin{equation}
  \big\langle \ell_{\rm 2D} \big\rangle
  \;=\; \frac{\pi}{4}\,\big\langle \ell_{\rm 3D} \big\rangle
  \;\Rightarrow\;
  \big\langle \ell_{\rm 3D} \big\rangle
  \;\approx\; \frac{4}{\pi}\,\big\langle \ell_{\rm 2D} \big\rangle
  \approx 1.27\,\big\langle \ell_{\rm 2D} \big\rangle.
  \label{eq:pi4_correction}
\end{equation}

The same scaling applies to any statistic that is linear in separations, such as
the mean NN distance or the total edge length of a minimum
spanning tree (MST).
However, Eq.~\eqref{eq:pi4_correction} represents the expected value of the projection factor: Individual
separations exhibit a large scatter because $\sin\theta$ varies from zero to one.
For an isotropic population, one finds
$\langle \sin^{2}\theta\rangle = 2/3$ and
$\mathrm{Var}(\sin\theta)\approx 0.05$, implying a one--sigma spread
$\sigma_{\sin\theta}\approx 0.22$ in the projection factor.
Departures from isotropy (e.g.\ flattened, filamentary, or fractal geometries), selection effects, and boundary truncation
can further broaden the distribution. These caveats motivate validating the
$\pi/4$ correction against controlled numerical models covering a range of
radial profiles and morphologies.

The analysis above treats each pair in isolation, but
real core catalogues contain many objects whose collective spatial statistics
introduce an additional scaling with sample size.
Even without projection, the mean NN length depends systematically on both the
dimensionality of the problem and the number of points.
For a uniform Poisson process, the characteristic spacing simply reflects the
typical inter-point distance, which is
\(\langle\ell_{\rm 2D}\rangle \propto (A/N)^{1/2}\) in 2D and
\(\langle\ell_{\rm 3D}\rangle \propto (V/N)^{1/3}\) in 3D
\citep{Clark1954, Casertano1985}.
Thus the ratio scales as \citep[see also][]{Schmeja2005}
\begin{equation}
  \frac{\langle\ell_{\rm 3D}\rangle}{\langle\ell_{\rm 2D}\rangle}
  \;\propto\; N^{1/6},
\end{equation}
increasing slowly with the number of objects.
Within a unit sphere, the ratio is about $\sim1.5$ at $N=10$ and $\sim2.2$ at
$N=100$.
Hence crowding, or equivalently the number of objects per volume, also drives
the effective 3D--to--2D ratio upward, meaning a single global factor of
$4/\pi\simeq1.27$ cannot capture this behaviour.

Several alternative strategies have been proposed to relate projected fragment
separations to their intrinsic three-dimensional spacing.
\citet{Myers2017} derived analytic expressions for the mean fragment spacing in
spherical star-forming zones, treating each as fragmenting into multiple
“minimum unstable spheres”. He defined the intrinsic spacing as the cube root
of the volume-per-fragment and related it to the projected spacing through the
global radius, $R$:
\begin{equation}
  \ell_{\rm 3D} \;\approx\; \ell_{\rm 2D}\,
  \left(\frac{4R}{3\,\ell_{\rm 2D}}\right)^{1/3}.
  \label{eq:myers_spherical}
\end{equation}
This expression makes explicit that the mapping between projected and intrinsic
spacings depends on global geometry rather than a fixed $\pi/4$ factor.

Building on this geometric framework, \citet{Svoboda2019} and
\citet{Traficante2023} extended the analytic expression derived by
\citet{Myers2017} for a spherical ensemble
(Equation~\ref{eq:myers_spherical}). They predicted a mean 3D-to-2D correction
factor of $\simeq1.84$. However, when performing Monte Carlo sampling of the
unknown line-of-sight positions consistent with the observed 2D configuration,
they obtained a smaller empirical factor of $\simeq1.42$. This discrepancy further 
demonstrates that projection bias depends sensitively on the underlying spatial
structure and sampling of the core population.

In this paper we present a systematic numerical calibration of projection effects on NN statistics. 
Using controlled ensembles of spherical and hierarchical fractal models, we compare intrinsic 3D and projected 2D separations across a broad range of sample sizes, morphologies, and resolutions (Section~\ref{sec:results_sym}). 
We begin with a uniform baseline to establish the simplest geometric behaviour (Section~\ref{sec:nn_uniform}), quantify its scaling with sample size (Section~\ref{sec:N_dependence}), and then introduce the effects of finite resolution and beam blending (Section~\ref{sec:dynrange}). 
The influence of anisotropy and fractal geometry is examined in Section~\ref{sec:fractal_orient}, and the combined dependence on structure, sampling, and resolution is synthesised into an empirical projection–correction function $\mathcal{C}(N,\mathrm{SDR})$ in Section~\ref{sec:projection_scaling}. 
Finally, in Sect.~\ref{sec_disc} we discuss the implications for interpreting observations and simulations, outline caveats, and summarise a practical prescription for converting observed 2D core separations into physically consistent 3D values.
The \texttt{corespacing3d} package used to produce the analysis presented in this
work, and to apply the $\mathcal{C}(N,\mathrm{SDR})$ correction factor, is publicly
available online \citep{Barnes2026}.\footnote{https://zenodo.org/records/18195695}

\section{Model framework}
\label{sec:models}

To quantify the impact of projection on measured core separations, we constructed a
set of simple three-dimensional toy models for the spatial distribution of dense
cores. These models allowed us to compare the true intrinsic separations to their
two-dimensional projected counterparts under controlled conditions, and to assess
the statistical validity of the $4/\pi$ correction factor introduced above.

\subsection{Core placement}

We populate a region of characteristic radius $R$ with $N$ point-like ``cores''.
We consider two families of models:

\smallskip

(i) Spherical profiles.
Isotropic distributions within a sphere of radius \(R\), including a uniform
baseline and several centrally concentrated profiles (Gaussian, power law,
Plummer; see Appendix~\ref{app:profiles}). In the main text we focus on the
uniform case, which provides a transparent baseline; the other profiles show
very similar trends with modest shifts in absolute scale.

\smallskip
(ii) Hierarchical fractals (anisotropic).
To capture clumpy, filamentary substructure, we generated fractal point sets
following \citet{Cartwright2004}. A cube is recursively subdivided into
$n_{\rm div}^3$ cells; at each generation, cells are retained with probability
$p_{\rm keep}=n_{\rm div}^{\,D-3}$, where $D$ is the fractal dimension
($0<D\le3$). The recursion proceeds until the number of occupied cells exceeds
the desired sampling size $N$ by a comfortable margin (i.e.\ until the structure
reaches sufficient multiplicity to populate the target number of points after
pruning). We then optionally restrict the distribution to the inscribed sphere of
radius $R$, rescale coordinates, and add small Gaussian jitter within final cells
to break grid regularity. Lower $D$ yields highly clumpy, anisotropic structures,
whereas $D\!\rightarrow\!3$ approaches uniformity. Because the resulting geometry
is not spherically symmetric, projected NN statistics can depend on viewing angle
(see Sect.~\ref{sec:fractal_orient}).

\subsection{Projection from 3D to 2D}
Once generated, the 3D distribution is rotated by a chosen set of
Euler angles $(\alpha,\beta,\gamma)$ and then projected onto the $x$--$y$ plane by
discarding the $z$ coordinate. This procedure mimics the observational situation
where only the sky-plane separations are accessible, while the line-of-sight
dimension is collapsed.  

\subsection{Nearest neighbour analysis}
For each configuration, we computed the NN graph in
three dimensions and in the two-dimensional projection. Each core is
linked to its single closest companion, producing a directed graph that can be
collapsed into a set of unique, undirected edges.
The output of the NN analysis consists of (i) the set of nearest-neighbour edges, i.e.~the pairs of nodes identified
as closest companions, and (ii) the length of each edge, corresponding to the minimum separation of
each core.

For context,
the MST is another pairwise graph used widely in
clustering studies. While the NN graph isolates local proximity (sensitive to
fragmentation scales), the MST enforces global connectivity and balances short
and long links, making it well suited to tracing overall geometry and
sub-clustering. We include MST-based comparisons alongside NN results in Appendix\,\ref{appendix:discussion_nn_mst} to
illustrate how projection affects global connectivity versus local spacing.

\section{Model results}
\label{sec:results_sym}

To isolate and understand the different factors that shape the observed 
NN statistics, we begin with the simplest possible case and 
then progressively add complexity. 
Starting from well sampled ($N=200$) uniform random distributions establishes the geometric baseline 
against which all subsequent effects can be measured. 
We then explore how finite sampling the distribution and stochastic variance affect the stability 
of the NN network (varying $N=2 - 200$), followed by how observational limitations alter its apparent topology. 
Finally, we examine a more physically motivated, hierarchical fractal distribution 
to assess how intrinsic substructure and orientation introduce additional, 
direction--dependent biases. 

\subsection{Uniform baseline: The simplest case}
\label{sec:nn_uniform}

\begin{figure*}[t]
    \centering
    \includegraphics[width=\linewidth]{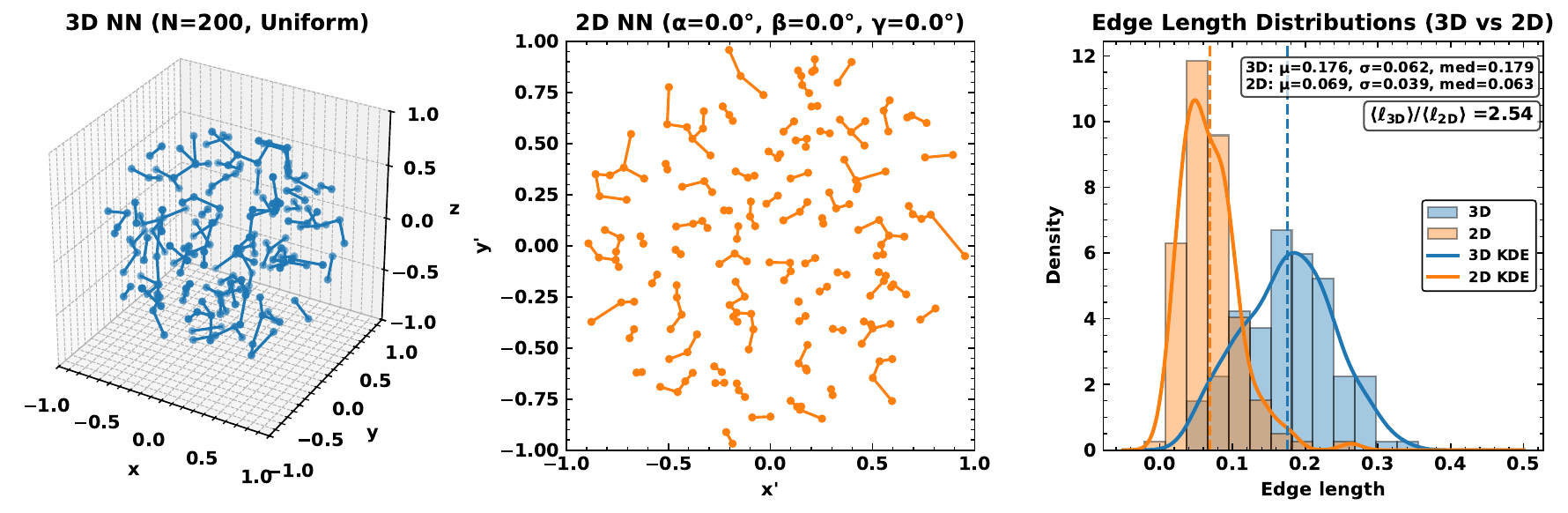} \\
    \includegraphics[width=\linewidth]{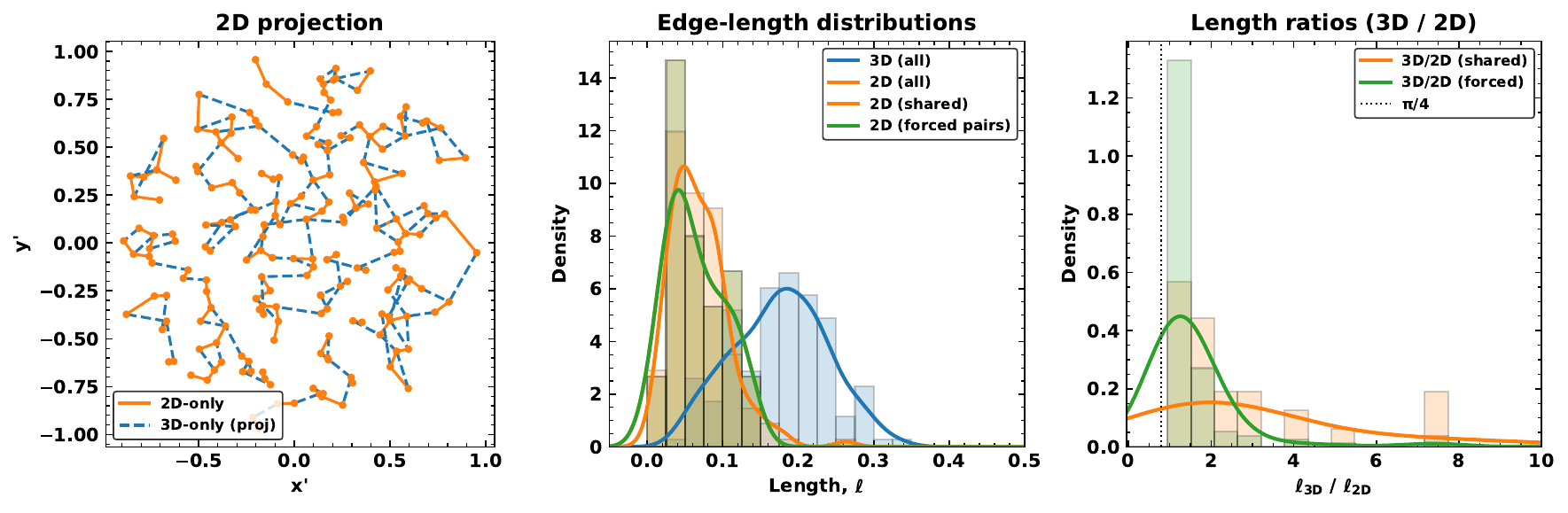}
    \caption{
        Nearest--neighbour comparisons for $N=200$ points drawn from a uniform spherical distribution
        within $R=1.0$. 
        Top: NN graph constructed in 3D, where the blue connections are the 3D NNs (left); 2D projection, where the orange connections are the 2D NNs (Centre); and the corresponding
        distributions of NN edge lengths (right) highlighting the overall shortening in projection.
        Bottom: Detailed breakdown. The left panel repeats the 2D overlay of 3D and 2D edges,
        while the right panels show the distributions of $\ell_{\rm 3D}$, $\ell_{\rm 2D}$ for shared edges
        and the ratios $\ell_{\rm 3D}/\ell_{\rm 2D}$. Vertical dashed lines mark the medians, and KDEs are overplotted on the histograms. Together these panels illustrate how
        projection both compresses lengths and reassigns neighbours, with only the very shortest pairs
        surviving unchanged.
    }
    \label{fig:nn_uniform}
\end{figure*}

We started with the simplest model: a uniform 3D distribution of $N=200$ points
within a sphere of radius $R=1$.  Figure~\ref{fig:nn_uniform} shows the NN graph in 3D
(top-left panel), the NN graph constructed from the projected points in 2D (top-middle panel), and
the corresponding distributions of NN edge lengths (top-right panel). Projection
immediately reduces the apparent neighbour separations to then give,\footnote{Throughout this
paper we define the correction factor as the ratio of ensemble
means, \(\mathcal{C}=\langle\ell_{\rm 3D}\rangle/\langle\ell_{\rm 2D}\rangle\), rather
than the mean of individual ratios
\(\langle\,\ell_{\rm 3D}/\ell_{\rm 2D}\,\rangle\). The two coincide only when
the same pairs are compared (e.g.\ in the `shared' or `forced' analyses).}
\[
\mathcal{C} = \frac{\langle\ell_{\rm 3D}\rangle}{\langle\ell_{\rm 2D}\rangle}
= 2.5 \quad (N=200),
\]
which is far larger than the simple geometric expectation of $4/\pi \simeq 1.27$.

The overlap between the 3D and projected 2D NN graphs is modest. Only about
20\% of the 3D links are retained after projection, and the Jaccard similarity ($J$)
is just 0.12 (the Jaccard similarity or index quantifies the fractional overlap
between two sets, $A$ and $B$, and is defined as \( J = |A \cap B| / |A \cup B| \). It ranges
from 0 for no shared elements, to 1 for identical sets). The few edges that survive in both graphs correspond mainly to the smallest
true separations and remain strongly compressed,
$\langle \ell_{\rm 3D}\rangle/\langle \ell_{\rm 2D} \rangle \simeq 2.0$,
well above the geometric baseline of $4/\pi \simeq 1.27$. 
Projection therefore shortens even the closest pairs substantially more than
simple foreshortening alone.

Longer 3D connections are typically broken and replaced after projection.  
In the bottom–left panel of Fig.~\ref{fig:nn_uniform}, the shared 3D-only edges that exist in the 3D NN graph but disappear in 2D
(shown in blue) span
relatively long intrinsic separations
($\langle \ell_{\rm 3D} \rangle_{\rm 3D\text{-only}} \simeq 0.18$), which project to
$\langle \ell_{\rm 2D} \rangle_{\rm proj(3D\text{-only})} \simeq 0.15$.\footnote{
“Shared” edges are pairs that remain NNs in both 3D and 2D after projection, 
tracing links that survive topologically.  
“Forced” edges use the same 3D pairs but measured in 2D projection, isolating pure geometric foreshortening.  
}  
These lost connections are almost always replaced in the projected NN graph by
2D-only edges (shown in orange in bottom–left panel of Fig.~\ref{fig:nn_uniform}), i.e.\ new neighbours that appear only after
projection. These are much shorter, with
$\langle \ell_{\rm 2D} \rangle_{\rm 2D\text{-only}} \simeq 0.07$.  
This behaviour is visible in the bottom–middle panel of
Fig.~\ref{fig:nn_uniform}, where the 2D-only histogram is concentrated at
small scales compared to the projected 3D-only distribution.  
Quantitatively, the reassigned edges lie on average
$\Delta\langle\ell\rangle\simeq -0.08$ below the projected lengths of the
removed 3D-only links, and $\sim\!96\%$ of 2D-only edges fall below the mean
projected length of their 3D predecessors.  
Together, these trends confirm that projection into 2D space not only
foreshortens individual separations but, more importantly, rewires the
NN network by replacing long intrinsic links with new, much shorter projected
neighbours, so that the apparent contraction of the 2D NN network is driven
predominantly by systematic neighbour reassignment rather than pure geometric
shortening.

As a complementary check, we can also ask what happens if the NN pairs
themselves are forced to be fixed. When the same 3D pairs are forced to be measured in projection, the mean ratio
drops slightly below the geometric value,
$\langle \ell_{\rm 3D} / \ell_{\rm 2D} \rangle_{\rm forced} = 1.1$,
indicating that pure foreshortening alone cannot reproduce the observed
$2.5\times$ contraction.  This behaviour is seen in the bottom–right panel of
Fig.~\ref{fig:nn_uniform}, where the distribution of $\ell_{\rm 3D}/\ell_{\rm 2D}$
for forced pairs (green) peaks near unity, while the shared-pair distribution
(orange) extends to much larger ratios, tracing the stronger compression produced
by reassignment.  At the node level, only about $8\%$ of points retain identical
neighbours in 3D and 2D, demonstrating that projection reorganises the local
topology of almost the entire network.

\subsection{Dependence on sample size}
\label{sec:N_dependence}

\begin{figure*}[ht!]
    \centering
    \includegraphics[width=\textwidth]{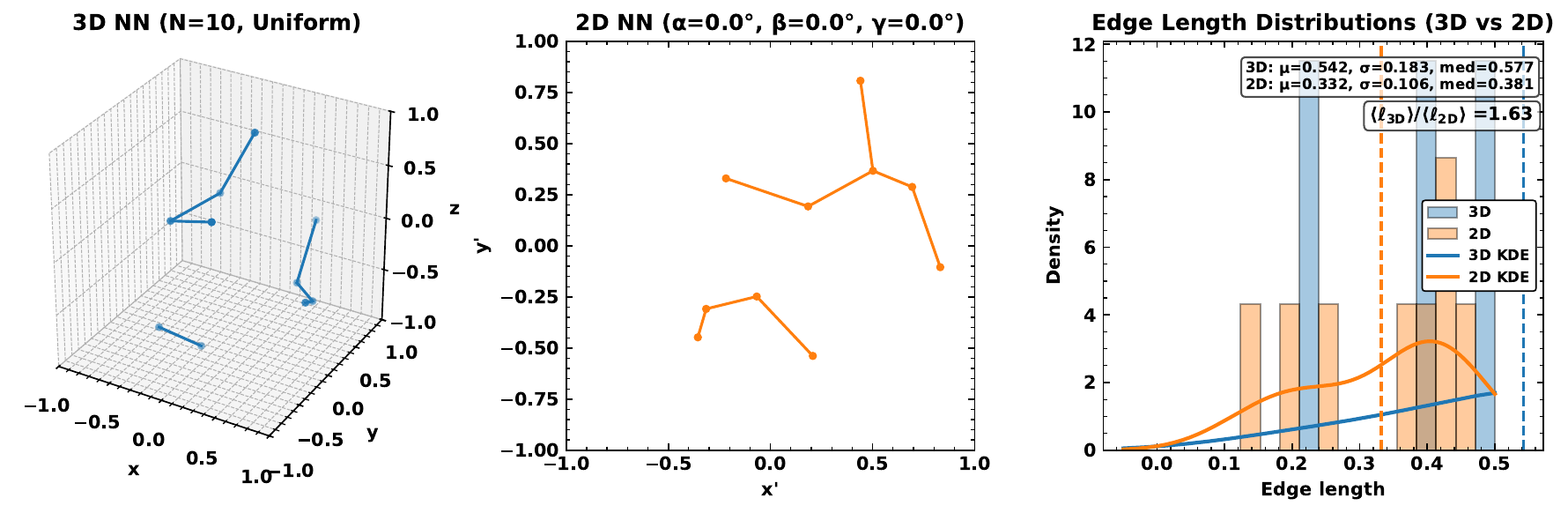}
    \includegraphics[width=\textwidth]{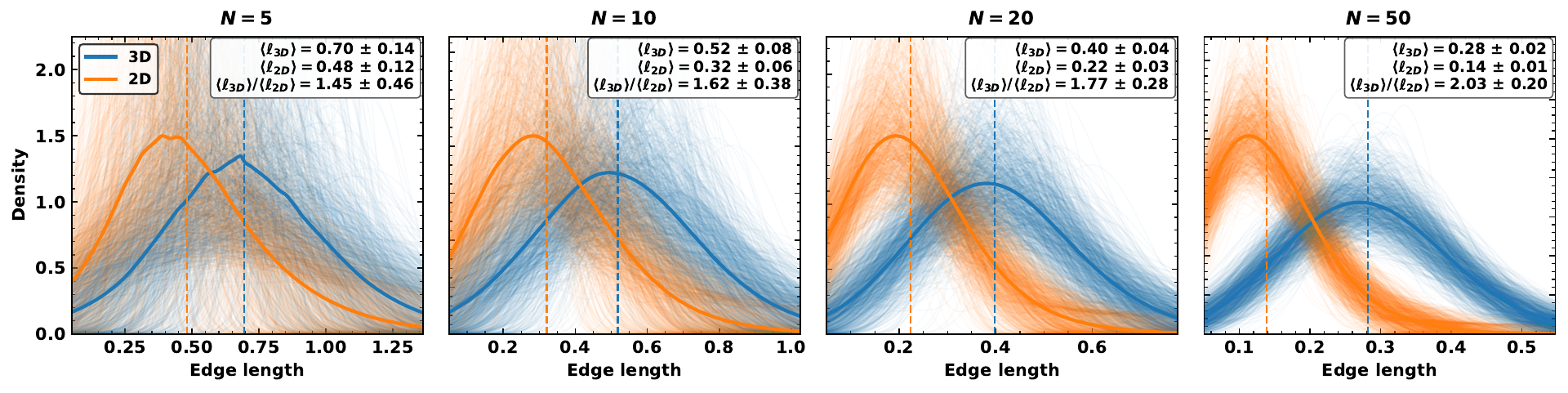}
    \caption{
        Nearest-neighbour comparisons for uniformly sampled point distributions. 
        Top: Example realisation with $N = 10$ points drawn within a sphere of radius of $R = 1.0$. 
        The NN graph constructed in 3D (left), its 2D projection obtained by dropping the line-of-sight coordinate (centre), 
        and the corresponding edge-length distributions (right) illustrate how projection systematically shortens apparent separations 
        and modifies the network connectivity even in a simple isotropic configuration. 
        Bottom: Results from the Monte Carlo ensembles ($10^3$ realisations each) for $N = 5$, 10, 20, and 50 
        showing stacked kernel density estimates of 3D (blue) and projected 2D (orange) NN edge lengths. 
        The distributions converge at a stable median ratio as $N$ increases, 
        while stochastic fluctuations dominate at small values of $N$.
    }
    \label{fig:nn_uniform_lowstats}
\end{figure*}

The structure of the NN network naturally depends on the
number of sampled cores. To illustrate its behaviour in the simplest case,
the upper panels of Fig.~\ref{fig:nn_uniform_lowstats} shows an example for a uniform random
distribution of $N = 10$ points within a sphere of radius $R = 1.0$.
Even in such a statistically isotropic configuration, the 2D projection shortens 
the apparent edge lengths (median ratio $\ell_{\mathrm{3D}}/\ell_{\mathrm{2D}}\!\simeq\!1.6$, 
closer to the geometric expectation of $4/\pi$) and modifies the network topology: 
only about $57\%$ of the 3D connections are recovered in projection 
($J = 0.36$). Several intrinsic 3D links are replaced by shorter projected ones, 
reflecting how chance alignments and foreshortening affect the apparent 
connectivity. The example also highlights the stochasticity inherent to 
small--$N$ realisations, where the precise NN configuration and the resulting 
length ratios vary between random draws of an identical underlying distribution.

The lower panels of Fig.~\ref{fig:nn_uniform_lowstats} quantify how the
projection bias depends on sample size for the uniform case using Monte Carlo
ensembles ($10^3$ realisations per $N$). For each run we computed the ratio
$\mathcal{C}(N) \equiv \langle \ell_{\mathrm{3D}} \rangle / \langle \ell_{\mathrm{2D}}
\rangle$ and then characterised its distribution across the ensemble. The
median ratio increases systematically with $N$, in line with the expected
$N^{1/6}$ scaling for Poisson samples (reflecting $\langle
\ell\rangle\!\propto\!N^{-1/3}$ in 3D and $N^{-1/2}$ in 2D). For example, we
find $\mathcal{C}(N{=}5) = 1.50 \pm 0.35$ (16–84\%: 1.20–1.79), $\mathcal{C}(N{=}10) = 1.65 \pm
0.28$ (1.39–1.91), $\mathcal{C}(N{=}20) = 1.80 \pm 0.23$ (1.58–1.99), and $\mathcal{C}(N{=}50) =
2.05 \pm 0.16$ (1.89–2.21), where the quoted uncertainties are the run–to–run
scatter in $\mathcal{C}$. As $N$ increases the KDE stacks in the lower panels become
smoother, the ensemble of $\mathcal{C}$ values narrows, and the 3D and 2D distributions
separate more cleanly. The latter is quantified by the Kolmogorov–Smirnov
statistic $D$, which measures the maximum difference between the cumulative
distributions: $D$ rises from $\simeq 0.3$ at $N{=}5$ to $\simeq 0.6$ at
$N{=}50$, indicating progressively stronger distinction between the 3D and 2D
spacing distributions. Small-$N$ realisations remain dominated by stochastic
scatter, but by $N\!\gtrsim\!20$ the median behaviour stabilises and the
ensemble converges towards the combined geometric–statistical expectation, with
$\mathcal{C}(N)$ well above the pure-projection baseline $4/\pi\simeq1.27$ due to the
different $N$–scalings of $\langle \ell\rangle$ in 3D and 2D.

\subsection{Impact of beam blending and spatial dynamic range}
\label{sec:dynrange}

\begin{figure*}[ht!]
    \centering
    \includegraphics[width=\textwidth]{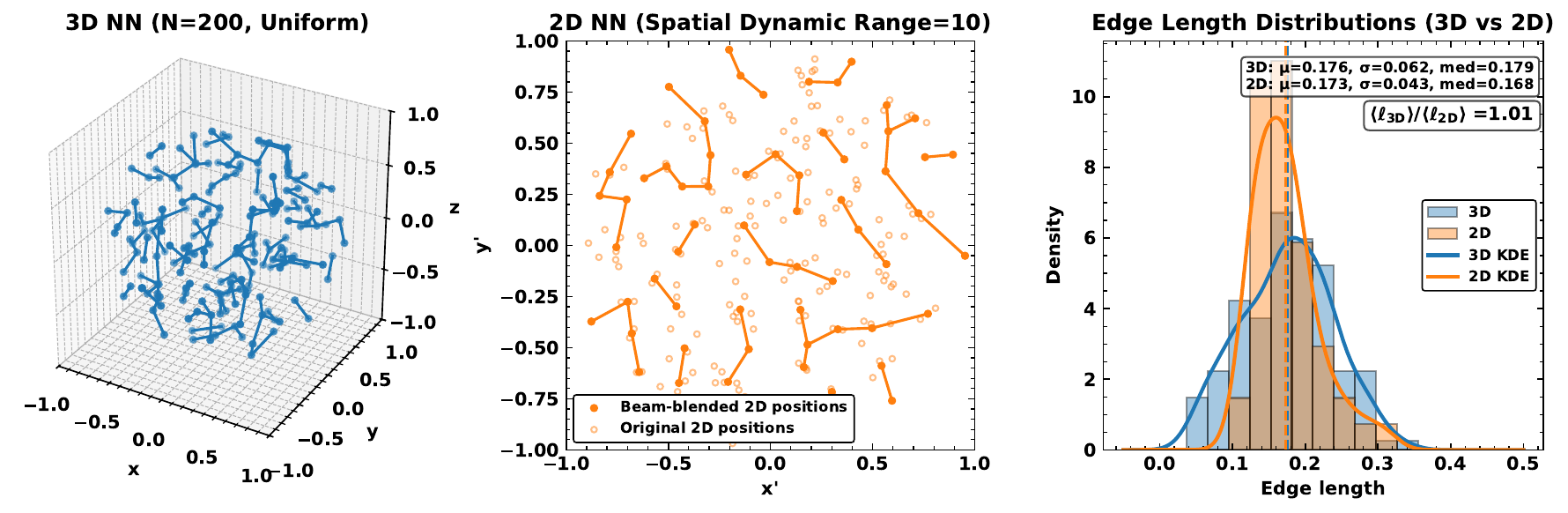}
    \caption{
        Illustration of the impact of a finite SDR on NN statistics
        using a uniform distribution of $N = 200$ points within a sphere of radius $R = 1.0$. 
        Left: Intrinsic 3D NN graph. 
        Centre: Two-dimensional projection after applying a beam--blending step that merges points closer than 
        one beam width, corresponding here to a SDR of $\mathrm{SDR} = \mathrm{FoV}/\mathrm{FWHM}_{\rm beam} = 10$. 
        Circles mark the original 2D positions (open) and the resulting beam--blended centroids (filled). 
        Right: Distributions of NN edge lengths in 3D and 2D after blending. 
        In this example, $135$ of the $200$ projected cores (67.5\%) are merged into $65$ effective groups, 
        erasing all topological correspondence between the intrinsic and projected networks 
        ($J = 0.00$, overlap fraction = 0). 
        The typical 3D and 2D NN lengths become nearly equal 
        ($\langle \ell_{\mathrm{3D}}\rangle / \langle \ell_{\mathrm{2D}}\rangle \simeq 1.0$), 
        as beam blending suppresses the shortest intrinsic separations 
        that normally produce the geometric compression factor 
        ($4/\pi \simeq 1.27$). 
        This example illustrates that limited spatial resolution can strongly distort the apparent connectivity 
        and scale distribution of dense cores even in an intrinsically uniform configuration.
    }
    \label{fig:nn_uniform_dynrange}
\end{figure*}

Real observations impose a finite spatial dynamic range (SDR) and, in practice, a limited ability to resolve close pairs. In this section we explicitly test how finite resolution and beam smoothing (i.e.\ confusion and blending) modify the NN network. For convenience, we express resolution through a single dimensionless parameter:
\[
\mathrm{SDR} \equiv \frac{\mathrm{FoV}}{\mathrm{FWHM}_{\rm beam}}\, ,
\]
which simply measures how many synthesised beams fit across the field of view (FoV). This quantity can be rescaled straightforwardly to any instrument and setup. For instance, in an ALMA observation at 100\,GHz, the primary beam has a full width half maximum (FWHM) of about 57\arcsec, so a single pointing imaged at 1\arcsec\ resolution would have $\mathrm{SDR}\!\approx\!57$ (mosaicked observations would have a larger FoV).\footnote{%
Here we distinguish between the FoV, which sets the number of beams across the image, and the maximum recoverable scale (MRS) of an interferometric observation, which defines the largest angular structure that can be reliably imaged before spatial filtering removes extended emission. While the MRS is crucial for studying diffuse or filamentary envelopes, our analysis focuses on small-scale fragmentation within the primary beam, and specifically the spacing and number of compact cores that fall within the FoV. The relevant parameter is therefore the number of synthesised beams across the primary beam (\(\mathrm{SDR}\)), rather than the ratio of the smallest to largest recoverable scales, with the caveat that the MRS can still impact structures comparable to the beam size depending on the large-scale emission morphology (see \citealp{Plunkett2023}).}

When the $\mathrm{SDR}$ is small (i.e. large $\mathrm{FWHM}_{\rm beam}$ for fixed FoV), 
sources separated by less than a $\mathrm{FWHM}_{\rm beam}$ cannot be
distinguished as individual objects in projection and appear merged in the
observed map. To mimic this observational limitation, we applied a simple
beam–blending step after projecting the intrinsic 3D distribution:
all points within one beam radius of each other are grouped together
and replaced by a single blended source at their mean position. For the
uniform example shown in Fig.~\ref{fig:nn_uniform_dynrange},
we adopted $\mathrm{SDR}=10$, corresponding to an effective resolution of $0.1$
in our normalised units.

At $N=200$ within $\mathrm{FoV} = R=1$, introducing finite resolution through beam
blending dramatically alters the NN network.
For $\mathrm{SDR}=10$ (beam FWHM $=\mathrm{FoV}/10$), any projected points
closer than one beam are merged into a single centroid, mimicking the loss of
distinct sources in observations. In this configuration, 135 of the 200
projected positions (67.5\%) collapse into 65 blended groups, leaving only
$N_{\rm eff}=65$ nodes for the 2D NN graph.
This coarse–graining removes precisely the shortest intrinsic separations (i.e. the
pairs most likely to be genuine neighbours in 3D) and thus erases the
topological correspondence between the intrinsic and projected networks;
no 3D NN edges are recovered in 2D.
At the same time, reducing the number of nodes inflates the characteristic 2D
spacing. 

Quantitatively, the intrinsic 3D NN length distribution has
$\mathrm{median}=0.179$ (mean $0.176\pm0.062$),
whereas the unsmoothed 2D projection has lengths
$\mathrm{median}=0.065$ (mean $0.069\pm0.021$),
reflecting strong projection compression ($\mathcal{C}\simeq2.5$; Section~\ref{sec:nn_uniform}).
After imposing finite resolution with $\mathrm{SDR}=10$, the adopted
effective resolution scale is 
$1/10=\;0.1$,
and the beam–blended 2D distribution shifts to
$\mathrm{median}=0.168$ (mean $0.173\pm0.043$).
Because the resolution now exceeds the typical unsmoothed projected separations
($\langle \ell_{\rm 2D}\rangle_{\rm unsm.}\approx0.069>1$),
the beam systematically removes precisely the short NN links that
dominated the projected network, merging close pairs into single centroids and
truncating the short-end tail of the 2D NN distribution.
As a result, the usual geometric compression seen under ideal projection largely
disappears: the mean ratio collapses to $\mathcal{C}(\mathrm{SDR}=10) \;\simeq\;1.0$.
In the unsmoothed case, by contrast, the same configuration yielded
$\mathcal{C}\simeq2.5$, with only $\sim$20\% of 3D NN edges surviving in projection and the
rest replaced by much shorter 2D companions.
Finite resolution therefore flips the regime from projection-dominated
(shortening via neighbour reassignment) to resolution-dominated
(coarse-graining via blending): the apparent network becomes nearly
isometric while the intrinsic connectivity is erased.

\subsection{Fractal, anisotropic structure: Orientation matters}
\label{sec:fractal_orient}

\begin{figure*}[t]
    \centering
    \includegraphics[width=\linewidth]{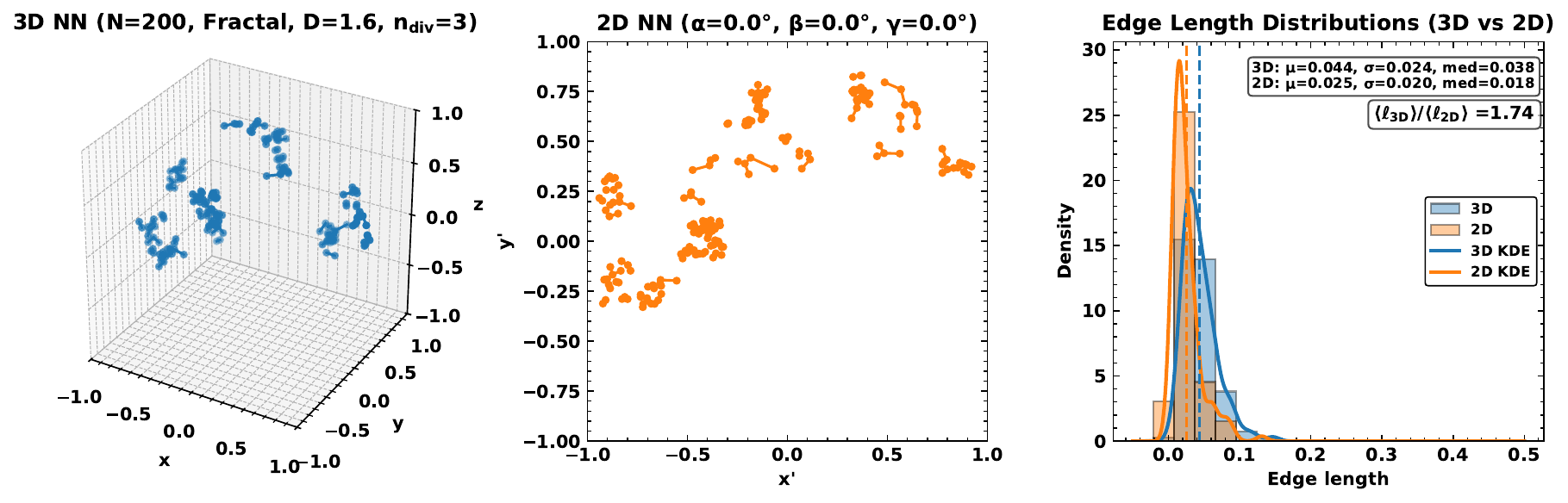} \\
    \includegraphics[width=\linewidth]{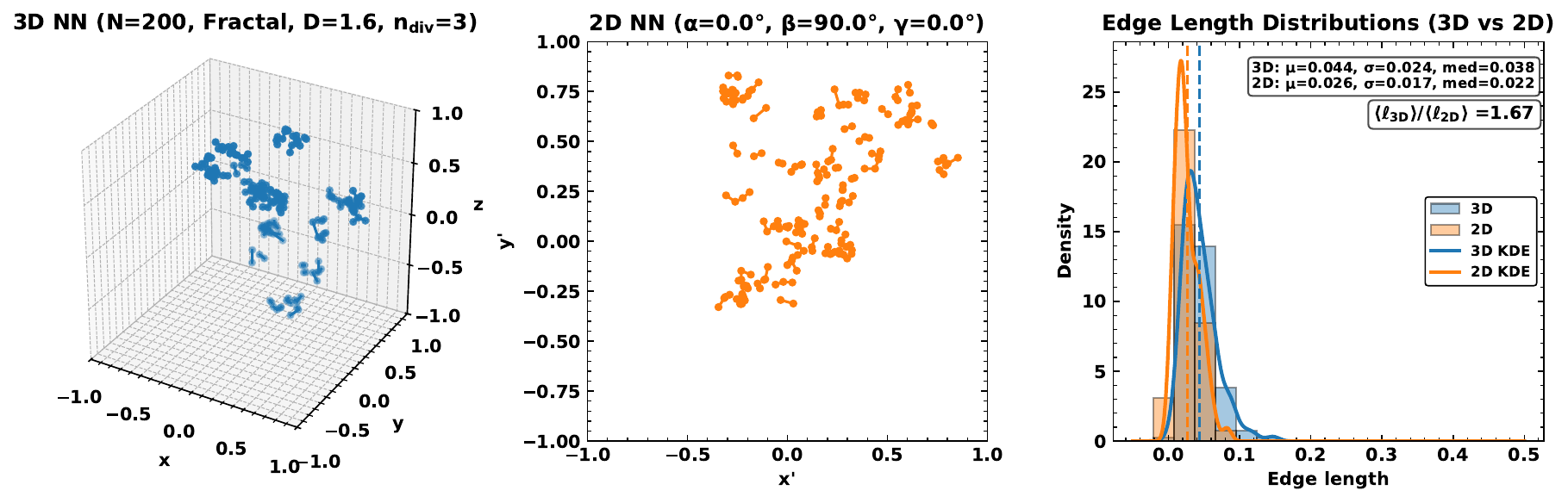} \\
    \includegraphics[width=\linewidth]{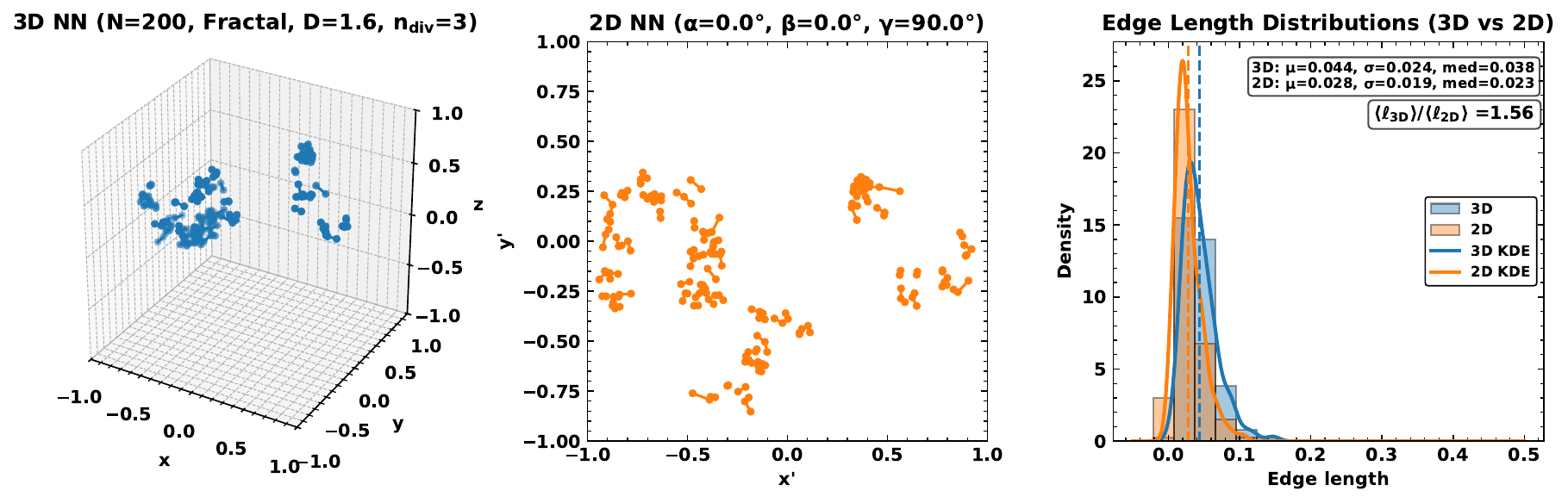}
    \caption{
        Nearest--neighbour graphs for a fractal distribution with
        $N=200$ points, fractal dimension $D=1.6$, and sub-division $n_{\rm div}=3$,
        within $R=1.0$. Each row shows the NN network in 3D (left),
        the projected NN network in 2D (centre), and the corresponding
        NN edge--length distributions (right).
        The three rows illustrate different viewing geometries:
        no rotation (top), rotation about the $y$--axis
        ($\beta=90^\circ$; middle), and rotation about the $z$--axis
        ($\gamma=90^\circ$; bottom). 
        Projection systematically shortens the apparent NN separations and
        rewires connectivity, but the degree of overlap and compression
        depends on the line of sight, reflecting the anisotropic and clumpy
        structure of the fractal distribution.
    }
    \label{fig:nn_fractal_orient}
\end{figure*}

To explore the impact of anisotropic, clumpy structure, we turn to a hierarchical
fractal model that lacks spherical symmetry and whose projected NN statistics
therefore depend on viewing angle. The distribution follows the
Cartwright–Whitworth prescription with fractal dimension \(D=1.6\) and
sub-division \(n_{\rm div}=3\) within a unit sphere (\(R=1\)). Each generation
divides space into \(n_{\rm div}^3\) cells and retains a fraction
\(n_{\rm div}^{\,D-3}\), producing a highly clumpy, cluster-like morphology
typical of young stellar regions (e.g.\citealp{Cartwright2004} infer Taurus has $D \sim$ 1.5), while remaining compact enough for robust
sampling at fixed \(N\). We initialise \(N{=}200\) points and compare three
orthogonal viewing angles (Fig.~\ref{fig:nn_fractal_orient}).

Projection again shortens apparent NN separations and rewires connectivity.
For the full 2D networks, the mean compression (expressed as a ratio of mean
lengths) spans
\[
\frac{\langle \ell_{\rm 3D} \rangle}{\langle \ell_{\rm 2D} \rangle}
\simeq 1.56\text{--}1.74
\]
across the three orientations (from mean $\ell_{\rm 3D}/\ell_{\rm 2D}=1.56,\,1.67,\,1.74$).
The overlap between the 3D and 2D NN edge sets is moderate to high for a fractal.
The fraction of 3D NN edges recovered in 2D is \(0.51\)–\(0.59\), and \(32\)–\(43\%\) of the nodes retain identical neighbour sets.

The pairs that survive in both graphs are intrinsically short and contract from
\(\langle \ell_{\rm 3D} \rangle\sim 0.033\)–\(0.036\) to
\(\langle \ell_{\rm 2D} \rangle\sim 0.022\)–\(0.023\), giving shared-edge ratios of
\[
\Big\langle \tfrac{\ell_{\rm 3D}}{\ell_{\rm 2D}} \Big\rangle_{\rm shared}
\simeq 1.30\text{--}1.41.
\]
By contrast, when the same 3D NN pairs are forced to be measured in
projection, the mean ratios sit near unity,
\[
\Big\langle \tfrac{\ell_{\rm 3D}}{\ell_{\rm 2D}} \Big\rangle_{\rm forced}
\simeq 0.87\text{--}0.97,
\]
which is consistent with a pure-geometry baseline for fixed pairs.

The gap between the “shared” and “forced” behaviours again points to
rewiring as the main driver of additional shortening: new 2D-only links
are systematically shorter (median \(\ell_{\rm 2D}\simeq 0.016\)–\(0.023\)) than the
projected lengths of the discarded 3D-only links (median \(\sim 0.035\)–\(0.040\)),
with \(75\%\)–\(87\%\) of 2D-only edges lying below the median projected scale of
their lost 3D counterparts. Orientation changes the exact values but not the basic picture: lower \(D\)
(more clumpy structure) produces slightly smaller global
\(\langle \ell_{\rm 3D}\rangle/\langle \ell_{\rm 2D}\rangle\) than smoother,
more uniform cases, but in all instances projection acts by both
geometrically foreshortening separations and rewiring the NN network
through neighbour reassignment.

\section{Putting it all together: Dependence on structure, sampling, and resolution}
\label{sec:projection_scaling}

\begin{figure}
\centering
\includegraphics[width=\linewidth]{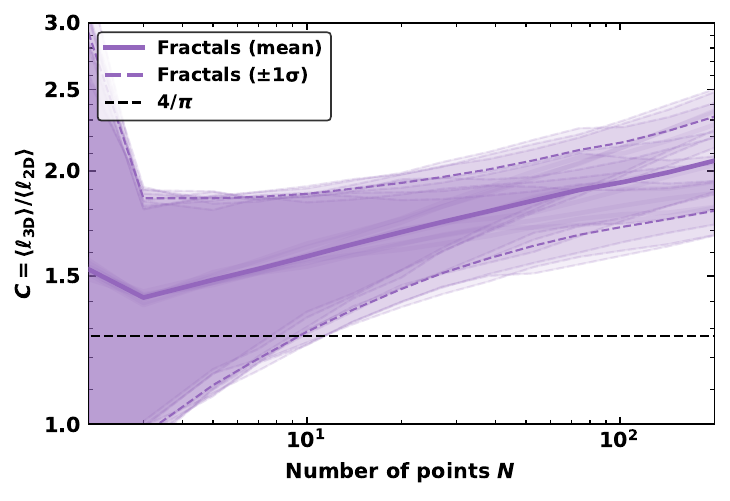}
\caption{
Nearest--neighbour projection ratio \(\ell_{\rm 3D}/\ell_{\rm 2D}\) versus sample size \(N\) for fractal
ensembles. For each \(N\), we averaged \(100\) realisations per configuration and then combined
across nine fractal setups \((D \in \{1.7,2.0,2.5\},\, n_{\rm div}\in\{2,3,4\})\) with
\texttt{jitter=True} and \texttt{prune=True}. The solid curve shows the cross-configuration mean,
and the dashed curves indicate \(\pm 1\sigma\) across configurations (not the run-to-run
uncertainty). The dashed horizontal line marks the geometric fixed-pair baseline \(4/\pi \simeq 1.27\).
Projection bias is present at all \(N\), with \(\ell_{\rm 3D}/\ell_{\rm 2D}\) increasing sub-linearly
with \(N\) and reaching \(\sim 2.0\)–\(2.3\) by \(N=200\), depending on fractal parameters.
A higher \(D\) (less clumpy) yields larger ratios at a fixed \(N\), indicating that beyond geometric
compression, crowding and neighbour reassignment drive additional shortening in projection.
}
\label{fig:ratio_vs_N_fractals}
\end{figure}

Having examined the effects of projection, sampling statistics, and resolution separately, 
we now bring these ingredients together to assess how the NN projection bias 
depends jointly on the intrinsic structure, the sample size $N$, and the effective SDR. 
Our goal is to establish a compact, quantitative description of the mean distortion factor 
\[
\mathcal{C}(N, \mathrm{SDR}) \equiv \frac{\langle \ell_{\mathrm{3D}} \rangle}{\langle \ell_{\mathrm{2D}} \rangle},
\]
and to determine how it varies across the physically relevant regimes for cluster and cloud observations.

We begin by isolating the dependence on $N$ alone. 
Figure~\ref{fig:ratio_vs_N_fractals} shows the ensemble-averaged projection ratio
for fractal realisations with dimensions $D=1.7$--$2.5$ and sub-division levels
$n_{\mathrm{div}}=2$--$4$, each averaged over 1000 random realisations.
The adopted range of fractal dimensions is intended to bracket a representative set of morphologies relevant for star-forming regions: lower values ($D\sim1.7$) produce highly filamentary, strongly clustered structures, whereas higher values ($D\sim2.5$) yield progressively more space-filling hierarchies. 
This choice is motivated by both numerical experiments of supersonic (approximately isothermal) turbulence and observational characterisations of molecular-cloud structure, which typically imply fractal dimensions in the broad range $D\sim1.8$--2.7, with systematic variations depending on the tracer, analysis method, and the balance of solenoidal versus compressive driving \citep[e.g.][]{Elmegreen1996,Sanchez2005,Sanchez2007,Federrath2009,Elia2014,Rathborne2015}.

The solid curve in Figure~\ref{fig:ratio_vs_N_fractals} represents the mean across all configurations, 
while the dashed lines indicate the $\pm1\sigma$ scatter between configurations 
(not the run-to-run variance within a single setup). 
As in the uniform case (\S\ref{sec:nn_uniform}--\ref{sec:N_dependence}), 
the projection bias increases systematically with $N$, 
rising from $\langle \ell_{\mathrm{3D}} \rangle / \langle \ell_{\mathrm{2D}} \rangle \!\sim\! 1.3$ 
for very small samples to $\sim 2.0$--$2.3$ by $N\simeq200$. 
This behaviour follows a sub-linear power law consistent with 
the geometric scaling $\langle \ell \rangle \!\propto\! N^{-1/3}$ in 3D 
versus $N^{-1/2}$ in 2D, yielding an effective exponent $\beta \!\simeq\! 1/6$. 
Higher fractal dimensions (i.e.\ smoother, more uniform point sets) 
tend to produce slightly larger ratios at fixed $N$, 
indicating that small-scale crowding and neighbour reassignment 
in the more clumpy ($D\simeq1.7$) cases partially offset the pure geometric compression.

\begin{figure}
\centering
\includegraphics[width=\linewidth]{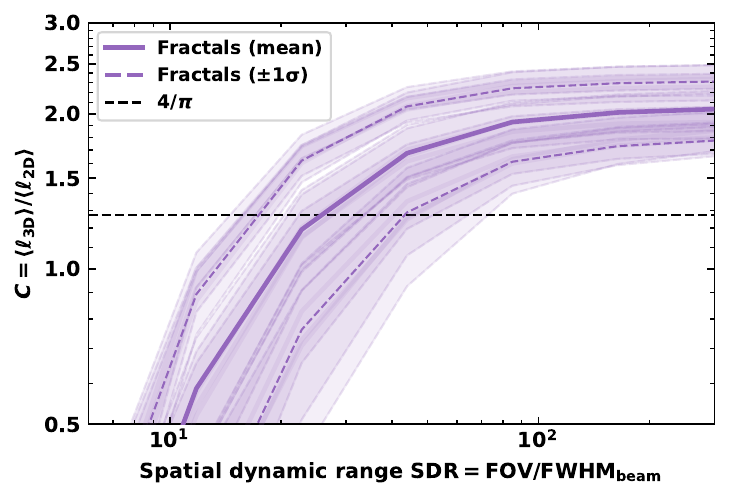}
\caption{
Nearest--neighbour projection ratio $\langle \ell_{\rm 3D} \rangle / \langle \ell_{\rm 2D} \rangle$
as a function of SDR evaluated for fractal ensembles 
($D=1.7$--$2.5$, $n_{\rm div}=2$--$4$) at a fixed value of $N=200$. 
Points and shaded bands show the mean and $\pm1\sigma$ scatter across configurations. 
The projection bias increases rapidly up to $\mathrm{SDR}\!\sim\!10$--$20$, 
beyond which it asymptotically approaches the intrinsic structural limit 
($\mathcal{C}_{\infty}\!\approx\!1.9$--$2.0$). 
At a low $\mathrm{SDR}$, beam blending dominates, merging close pairs and 
driving $\mathcal{C}\!\to\!1$, i.e.\ apparent isometry between 3D and 2D separations.
}
\label{fig:ratio_vs_DynRange_fractals}
\end{figure}

To disentangle the influence of spatial resolution, 
we next fix $N=200$ and vary the spatial dynamic range for  1000 fractal realisations with $D=1.7$--$2.5$, 
$n_{\mathrm{div}}=2$--$4$ 
(Figure~\ref{fig:ratio_vs_DynRange_fractals}). 
At high $\mathrm{SDR}$, the ratio saturates at $\mathcal{C}\!\approx\!2.0$, 
reflecting the intrinsic projection bias once individual cores are resolved. 
As $\mathrm{SDR}$ decreases, unresolved blending removes the shortest 
true separations, suppressing the apparent contraction until the 3D and 2D 
mean separations become comparable ($\mathcal{C}\!\approx\!1$). 
This transition occurs around $\mathrm{SDR}\!\sim\!20$ for the fractal models, 
but its exact location depends on the underlying spatial hierarchy: 
lower-$D$ distributions are affected at slightly higher $\mathrm{SDR}$ 
because their denser substructure contains more close pairs to be blended.
The dependence of the projection ratio on fractal dimension at fixed $(N,\mathrm{SDR})$, and the regime where structure and resolution become coupled, are examined in more detail in Appendix~\ref{app:ratio_vs_D}.

\begin{figure*}
\centering
\includegraphics[width=\linewidth]{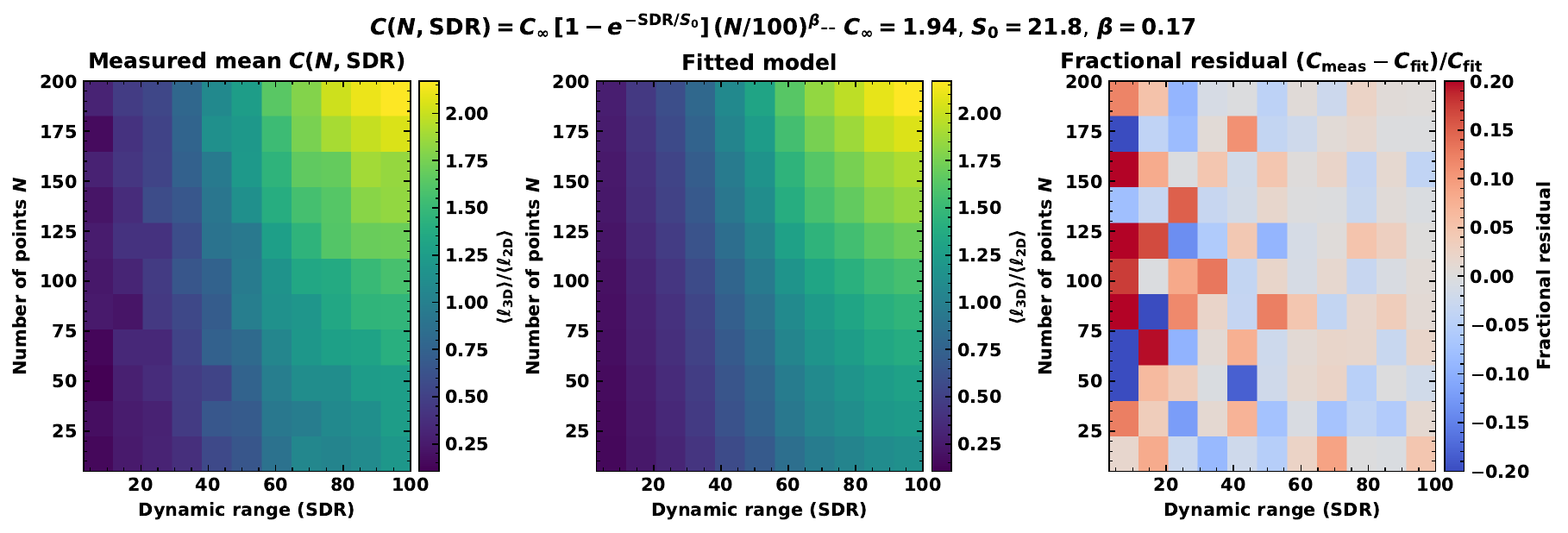}
\caption{
Combined dependence of the projection distortion ratio 
$\mathcal{C}(N,\mathrm{SDR}) = \langle \ell_{\rm 3D} \rangle / \langle \ell_{\rm 2D} \rangle$ 
on sample size $N$ and SDR. 
Left: Ensemble-averaged measurements from 1000 fractal realisations with $D=1.7$--$2.5$ and 
$n_{\mathrm{div}}=2$--$4$. 
Centre: Best-fitting three-parameter model 
$\mathcal{C}_\infty[1-e^{-\mathrm{SDR}/S_0}](N/100)^{\beta}$. 
Right: Fractional residuals $(\mathcal{C}_{\mathrm{meas}}-\mathcal{C}_{\mathrm{fit}})/\mathcal{C}_{\mathrm{fit}}$. 
The fitted parameters ($\mathcal{C}_\infty=1.94$, $S_0=21.8$, $\beta=0.17$) 
capture the joint behaviour across two orders of magnitude in both variables. 
The model converges to the geometric projection limit ($\mathcal{C}\simeq1.3$) 
for small well-resolved samples and to the beam-limited regime ($\mathcal{C}\simeq1$) 
when $\mathrm{SDR}\lesssim10$.
}
\label{fig:R_vs_N_SDR_simple_fit}
\end{figure*}

Finally, to capture the combined dependence of the projection ratio on both
sample size and spatial resolution, we modelled the full 2D surface
$\mathcal{C}(N,\mathrm{SDR})$ with a compact functional form that separates the scaling
with $N$ from the saturation with dynamic range:
\begin{equation}
\mathcal{C}(N,\mathrm{SDR}) \;=\;
\mathcal{C}_\infty\,\bigl[1 - \exp(-\mathrm{SDR}/S_0)\bigr]\,
\left(\frac{N}{100}\right)^{\beta}.
\label{eq:R_N_SDR}
\end{equation}
In this expression, $\mathcal{C}_\infty$ represents the asymptotic projection ratio for
large $N$ and effectively infinite resolution, $S_0$ defines the characteristic
dynamic range at which finite beam size begins to suppress small-scale
structure, and $\beta$ quantifies the weak residual scaling with $N$. Note here that, when applying this relation to observations, $N$ should be interpreted as the intrinsic three-dimensional population, such that using the observed $N$ yields a conservative lower limit; further practical caveats are discussed in Sect.~\ref{subsec:comp_obs}. Fitting this form to the grid of ensemble-averaged measurements from 1000 fractal realisations with $D=1.7$--$2.5$, 
$n_{\mathrm{div}}=2$--$4$ yields (see Figure~\ref{fig:R_vs_N_SDR_simple_fit})
\begin{align*}
\mathcal{C}_\infty &= 1.94 \pm 0.01,\\
S_0      &= 21.8 \pm 0.3,\\
\beta    &= 0.173 \pm 0.003.
\end{align*}
The $S_0$ term marks the onset of significant blending for data with
$\sim$10–20 independent resolution elements across the field. The analytic fit
reproduces the ensemble with a
fractional residual of $\sim 9$\,\%, indicating negligible systematic bias and
excellent agreement with the mean trends in both $N$ and $\mathrm{SDR}$.
To assess the intrinsic scatter around this mean relation, we separate
(i)~run--to--run Monte Carlo variation at fixed $(N,\mathrm{SDR})$ from
(ii)~differences between fractal geometries.

The Monte Carlo scatter is modest, with a median fractional dispersion of
$\sim 0.17$ (16–84\,\% range $0.10$–$0.26$).
In contrast, geometric variations introduce a substantially larger spread:
the median absolute deviation from the fitted surface is
$\lvert\Delta \mathcal{C}\rvert/\mathcal{C} \simeq 0.35$ (16–84\,\% range $0.12$–$1.19$), and the
full grid exhibits rare but extreme excursions up to
$\max\lvert\Delta \mathcal{C}\rvert/\mathcal{C} \simeq 4.5$. 
In practical terms, while the mean relation $\mathcal{C}(N,\mathrm{SDR})$ is
constrained to better than $\sim 10$\,\%, individual realisations of a fractal
point distribution can differ from this mean by $\sim 30-40$\,\% for typical
geometries, occasionally reaching $\sim 100$\,\% deviations, and in the most
extreme sparse or strongly hierarchical cases, by factors of several
(up to $\sim 450$\,\%; see Figure~\ref{fig:ratio_vs_N_fractals}).

In the limit of fully resolved observations 
($\mathrm{SDR}\!\rightarrow\!\infty$), the exponential term in 
Equation~\ref{eq:R_N_SDR} approaches unity, 
$[1-\exp(-\mathrm{SDR}/S_0)] \simeq 1$, 
and the relation simplifies to a purely sampling–dependent form:
\begin{equation}
\mathcal{C}(N) \;=\; 1.94\,\left(\frac{N}{100}\right)^{0.173}.
\label{eq:R_N_only}
\end{equation}
This expression captures the fundamental scaling of projection bias 
with sample size alone, independent of resolution effects, 
and reproduces the behaviour derived earlier for the idealised case ($N^{1/6}$; see Sect.~\ref{sec_int}).

\section{Discussion}
\label{sec_disc}

Projection alters NN statistics by both foreshortening
true separations and rewiring the NN graph. Many intrinsic 3D neighbours
are replaced by closer projected companions, so
\(\langle \ell_{\rm 3D} \rangle / \langle \ell_{\rm 2D} \rangle\)
systematically exceeds the geometric baseline \(4/\pi \simeq 1.27\) and grows
sub–linearly with sample size. Finite angular resolution adds a competing
effect: When the spatial dynamic range
\({\rm SDR}\equiv{\rm FoV}/{\rm FWHM}_{\rm beam}\) is modest, beam blending
merges sources within roughly one beam, removing the shortest links and
inflating the apparent 2D separations so that \(\mathcal{C}\) is driven back towards
unity as the characteristic NN scale approaches the beam. These trends are
captured by our empirical surface (Eq.~\ref{eq:R_N_SDR}), where
\(\mathcal{C}_\infty=1.94\) is the large–\(N\), infinite–resolution plateau,
\(\beta\simeq0.173\) matches the expected \(N^{1/6}\) scaling, and
\(S_0\simeq21.8\) marks the dynamic–range threshold where blending becomes
important.

Table~\ref{tab:R_vs_N_SDR} turns the fitted surface $\mathcal{C}(N,\mathrm{SDR})$ into a
set of ready–to–use numbers, organised by sampling ($N$) and SDR. Three regimes are evident.  

\begin{itemize}
    \item (i) Low $N$ or coarse resolution ($N\lesssim10$ or $\mathrm{SDR}\lesssim10$): Beam blending removes the shortest intrinsic links, and $\mathcal{C}$ sits near unity, so geometric de-projection alone is inadequate. 
    \item (ii) Intermediate ($N\sim20$–$100$, $\mathrm{SDR}\sim20$–$50$): $\mathcal{C}$ rises quickly as projection–induced rewiring dominates, while the resolution already resolves most pairs. This is the transition zone where comparing maps of different angular resolutions without correction can bias conclusions.  
    \item (iii) Well–sampled, well–resolved ($N\gtrsim100$, $\mathrm{SDR}\gtrsim50$): $\mathcal{C}$ approaches the large–$N$ plateau $\simeq2$, with weak additional gains from increasing $N$ or $\mathrm{SDR}$.
\end{itemize}

The second and third header rows in the table map $\mathrm{SDR}$ to the beam FWHM (for an ALMA
primary beam of $57\arcsec$ at 100\,GHz) and to a physical resolution for a
1\,pc field at 3.6\,kpc, allowing observers to locate their datasets in the
table directly. In practice, one chooses the row matching the $N$ and the column
closest to the $\mathrm{SDR}$ (or beam FWHM divided by physical resolution) and applies
$\mathcal{C}$ as a multiplicative correction to convert measured 2D NN spacings to an
approximate 3D mean. Differences on the order of 40\% between morphologies and
orientations set a sensible systematic floor for
error budgets. Within that tolerance, the tabulated values provide a concise,
physically motivated correction across typical ALMA setups.

\begin{table}[t]
\centering
\caption{
Fitted projection–correction factors \(\mathcal{C}(N,\mathrm{SDR})\) from the empirical relation
\(\mathcal{C}(N,\mathrm{SDR}) = \mathcal{C}_\infty [1 - e^{-\mathrm{SDR}/S_0}] (N/100)^{\beta}\)
using \(\mathcal{C}_\infty = 1.94\), \(S_0 = 21.8\), and \(\beta = 0.173\).
}
\label{tab:R_vs_N_SDR}
\begin{tabular}{lccccccc}
\hline\hline
\multicolumn{1}{c}{$N$} &
\multicolumn{7}{c}{$\mathrm{SDR}$} \\[2pt]
 & 5 & 10 & 20 & 50 & 100 & 200 & $\infty$ \\[3pt]
 & \multicolumn{7}{c}{$\mathrm{FWHM}_{\rm beam}$ [arcsec] (ALMA @ 100\,GHz)} \\[2pt]
 & 11.4 & 5.7 & 2.85 & 1.14 & 0.57 & 0.29 & 0 \\
 & \multicolumn{7}{c}{Physical resolution [pc] at 3.6\,kpc} \\[2pt]
 & 0.20 & 0.10 & 0.05 & 0.02 & 0.010 & 0.005 & 0 \\
\hline
5 & 0.24 & 0.43 & 0.69 & 1.04 & 1.14 & 1.16 & 1.16 \\
10 & 0.27 & 0.48 & 0.78 & 1.17 & 1.29 & 1.30 & 1.30 \\
20 & 0.30 & 0.54 & 0.88 & 1.32 & 1.45 & 1.47 & 1.47 \\
50 & 0.35 & 0.63 & 1.03 & 1.55 & 1.70 & 1.72 & 1.72 \\
100 & 0.40 & 0.71 & 1.16 & 1.74 & 1.92 & 1.94 & 1.94 \\
200 & 0.45 & 0.80 & 1.31 & 1.97 & 2.16 & 2.19 & 2.19 \\
\hline
\end{tabular}
\vspace{0.3em}

\tablefoot{
Values represent the mean ratio of intrinsic 3D to projected 2D NN
separations for a uniform distribution under finite spatial dynamic range
(\(\mathrm{SDR} = \mathrm{FoV}/\mathrm{FWHM}_{\rm beam}\)).
The tabulated values can be reproduced using the \texttt{projection\_correction(N, SDR)} function from the \texttt{corespacing3d} package \citep{Barnes2026}.
The second and third rows list the corresponding beam FWHM in arcseconds and
physical resolution in parsecs, assuming an ALMA primary beam of 57\arcsec\
at 100\,GHz and a typical clump diameter of 1\,pc (distance 3.6\,kpc).
For large samples (\(N \gtrsim 100\)) and well–resolved maps
(\(\mathrm{SDR} \gtrsim 30\)), the ratio saturates near \(\mathcal{C} \approx 2.0\),
whereas poorly resolved or low–$N$ cases converge to \(\mathcal{C} \approx 1\),
approaching the geometric limit \(4/\pi \simeq 1.27\).
For reference, an ALMA 100\,GHz pointing imaged at 1\arcsec\ resolution corresponds to
\(\mathrm{SDR} \approx 57\), equivalent to a spatial resolution of
\(\sim0.018\) pc for a 1 pc field at 3.6 kpc.
The calibration is based on idealised, complete simulations; when applied to real data with finite sensitivity, blending, and \(N_{\rm obs} < N\), these values
should be treated as approximate corrections and, if evaluated with \(N_{\rm obs}\),
as lower limits on the true 3D spacing (see Sect.~\ref{subsec_caveats}). }
\end{table}

\subsection{Comparison to observations}
\label{subsec:comp_obs}

\begin{figure}[t]
  \centering
  \includegraphics[width=\columnwidth]{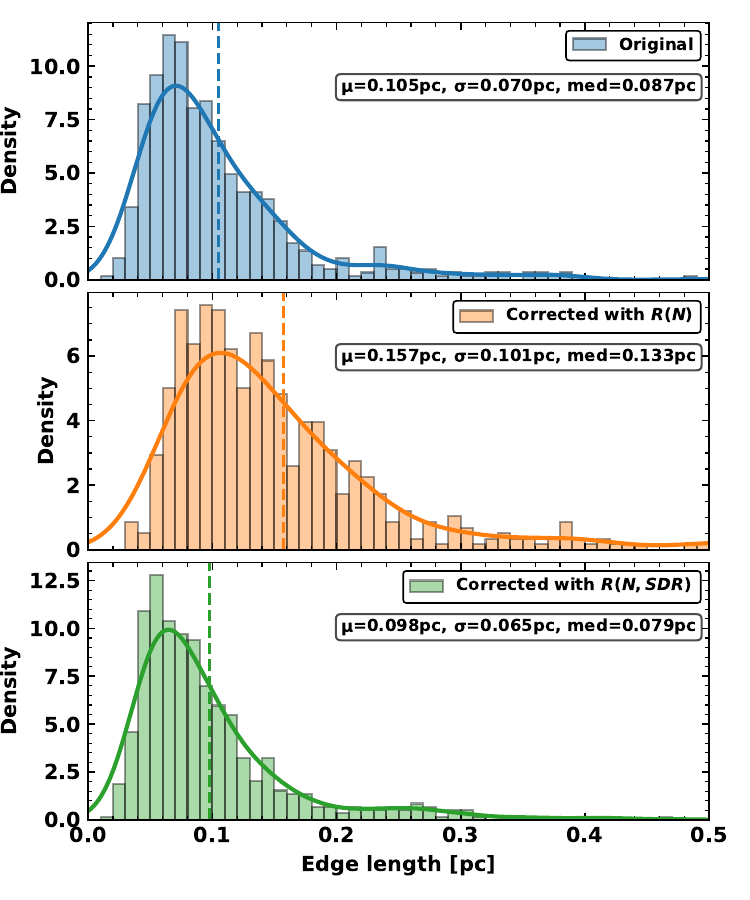}
  \caption{
  Effect of projection and resolution corrections on the distribution of NN 
  separations in the ASHES survey \citep{Morii2024}. 
  Top: Observed (projected) 2D separations.  
  Middle: Separations corrected for projection bias using the 
  sample–size–dependent relation $\mathcal{C}(N)$ (Eq.~\ref{eq:R_N_only}).  
  Bottom: Separations corrected for both projection and finite angular resolution 
  using $\mathcal{C}(N,\mathrm{SDR})$ (Eq.~\ref{eq:R_N_SDR}).  
  Dashed lines mark the median values in each panel.  
  The uncorrected distribution peaks near $0.08$\,pc, increases to $0.12$\,pc after applying $\mathcal{C}(N)$, 
  but returns to $0.07$\,pc when the finite dynamic range 
  ($\mathrm{SDR}\!\sim\!10$–20, defined as clump radius to beam FWHM) is included.  
  This demonstrates the competing influences of geometric projection, which lengthens 
  apparent separations, and beam blending, which suppresses the shortest scales (note the caveats of applying this correction to observations in Sect.~\ref{subsec_caveats}).  
  }
  \label{fig:obs_ashes_hist}
\end{figure}

To quantify how the choice of correction factor influences the inferred fragmentation scale, 
we applied the derived relations to the ASHES sample \citep{Morii2024}, 
which contains 839 dense cores across 39 IRDCs ($\sim$20 cores per region; $N$ varies from 8 to 39).  
For each region, we estimated the implied 3D NN separations using both 
the sample–size–dependent correction given by Eq.~\ref{eq:R_N_only} 
and the resolution–dependent form in Eq.~\ref{eq:R_N_SDR}, 
where the spatial dynamic range $\mathrm{SDR}$ is defined as the ratio of host clump radius (see \citealp{Morii2024}) to beam FWHM ($\sim$\,1\arcsec), and varies between 10 and 40.  
The resulting distributions are shown in Figure~\ref{fig:obs_ashes_hist}.

The statistics illustrate how each correction systematically alters the spacing distribution.  
The uncorrected (projected) separations have a mean of $0.096$\,pc and a median of $0.077$\,pc.  
Applying the $\mathcal{C}(N)$ correction increases these to $0.14$\,pc and $0.12$\,pc, respectively, 
broadening the distribution and shifting the median upward by roughly 50\%.  
This corresponds to a mean ratio of $1.50$, consistent with the ensemble prediction for $N\!\sim\!20$.  
In contrast, incorporating the finite dynamic range via $\mathcal{C}(N,\mathrm{SDR})$ lowers the mean and median 
back to $0.089$ and $0.073$\,pc, respectively -- effectively reducing the inferred 3D separations 
by $\sim$40\% relative to the $\mathcal{C}(N)$ case.  
This reversal arises because, for the ASHES data, the beam size is comparable to the typical NN spacing 
($\mathrm{SDR}\!\sim\!10$–20). Beam blending therefore suppresses the shortest separations 
and counteracts the geometric expansion produced by projection correction.

The implications of this result are subtle but important.  
In \citet{Morii2024}, the observed (projected) core separations were found to be comparable to 
the thermal Jeans length of the parent clumps, leading to the interpretation that fragmentation 
proceeds predominantly at the thermal, rather than turbulent, Jeans scale.  
Applying the $\mathcal{C}(N)$ correction increases the median spacing from $0.077$\,pc to $0.12$\,pc,
raising the ratio $\delta_{\rm sep}/\lambda_{\rm J}$ from $0.7$ to $\sim1.2$.  
This suggests that the intrinsic 3D separations are slightly larger than the thermal Jeans length, which is still 
consistent with thermally regulated fragmentation, but with less room for turbulent support.  
When finite angular resolution is included through $\mathcal{C}(N,\mathrm{SDR})$, the median spacing shifts back 
towards the uncorrected value ($0.073$\,pc), maintaining $\delta_{\rm sep}/\lambda_{\rm J}\!\simeq\!0.9$.  
However, this apparent “agreement” arises for the wrong reason: 
the dynamic–range correction suppresses separations because beam blending merges 
unresolved substructure into fewer larger nodes.  
This highlights an important warning for interpreting core spacings in marginally resolved maps, 
when the typical NN separation approaches the beam FWHM, 
a substantial fraction of the small–scale structure may remain unresolved, 
leading to biased estimates of both the physical spacing and the inferred fragmentation mode.  
In such cases, apparent consistency with the Jeans length does not necessarily confirm 
thermal fragmentation, but may instead reflect the observational limits of resolution and completeness.

We stress that this application is illustrative. In particular, the corrections are evaluated using the observed number of cores and an effective clump radius, whereas the intrinsic 3D sample size, geometry, and completeness are not known. As a result, the quoted 3D spacings should be treated as approximate and, if $N_{\rm obs}<N_{\rm true}$, as lower limits on the true projection factor and intrinsic separation, with systematic uncertainties of at least 30–40\,\% from the underlying morphology (see Sect.~\ref{subsec_caveats}).

\subsection{Comparison to simulations}
\label{subsec_compsim}

In simulations of star formation, and the subsequent dynamical evolution of star-forming regions, the spatial distributions of the stars are often quantified using NN analyses \citep[e.g.][]{Klessen2001b,Schmeja2006,Schmeja2011,Parker2018} and MST analyses \citep{Allison2010,Kirk2014,Parker2014,Dominguez2017,Parker2024}. For the majority of measured diagnostics (e.g. $\mathcal{Q}$-parameter, or the $\Lambda_{\rm MSR}$ measure of mass segregation), the  simulations can simply be analysed in 2D in the same way an observed star-forming region is. Often, a perfunctory check is made in a different plane (e.g. $z-x$ rather than $x-y$) to establish that the measurement is also significant from a different viewing angle, but diagnostics such as $\mathcal{Q}$ and $\Lambda_{\rm MSR}$  were designed specifically for analysis of 2D data and no conversion between two and three dimensions is performed.

However, similar clump-finding, or cluster-finding, algorithms are applied to simulation data as are applied to observed data \citep[e.g.][]{Guszejnov2022}, and similar issues arise in identifying structures in 2D rather than the true underlying three-dimensional distribution. Again, simulations can be analysed in the same number of dimensions available to observers (usually two), but as we have shown the two-dimensional structures identified may not accurately reflect the underlying distribution. For example, \citet{Parker2018} apply a Friends of Friends algorithm (which is a NN based algorithm) to $N$-body simulation data and identify groups in three dimensions, but in different projections the groups they identified in the $x-y$ plane were not always apparent in two dimensions in the $z-x$ and $z-y$ planes.

\citet{Parker2018} identified a further issue that is often overlooked in clump or cluster-finding algorithms, namely that clumpy and sub-clustered distributions may be self-similar on all scales (e.g. fractal), in which case identifying groups via, for example, an MST or NN method can become meaningless. 

\citet{Gutermuth2009} used a technique for identifying sub-clusters and substructures by constructing a MST of a star-forming region, and then analysing the distribution of MST branch lengths. They fit two lines to the distribution; one fitting the smallest branch lengths, and the other fitting the largest branch lengths. Where these two lines intersect is taken to be an indication of where small-scale structure changes to large-scale structure, and all branch lengths longer than this are removed in the full MST, leaving multiple smaller groups. 

However, \citet{Parker2015} find that if the distribution of branch lengths is plotted on a logarithmic axis, there is no obvious break length, which is due to the fractal distribution being similar on all scales. If there is no true break in an observed distribution, i.e.\,\,if it is self-similar on all scales (e.g. fractal), then any groups identified by this (or a similar technique) may not have any physical significance. This illustrates the pitfalls of over-interpreting these data, even when all of the three-dimensional information is available, such as in simulations.

\subsection{Caveats and future avenues of research}
\label{subsec_caveats}

Our analysis was deliberately focused on idealised toy models in order to isolate the
geometric effects of projection. Several caveats are worth highlighting.

\begin{itemize}

    \item Point-like cores. 
    Real dense cores have finite sizes determined by both the instrumental beam
    and their intrinsic density structure. Beam convolution, blending, and spatial
    filtering inevitably modify the observed core distribution, while cores sit atop
    a complex, non-uniform background of cloud emission. This makes their detection
    and deblending sensitive to the local surface-brightness field and to the
    specific core-identification algorithm employed. In projection, such effects can
    cause multiple nearby cores in 3D to merge into a single 2D source or, conversely,
    to fragment spurious subcomponents from structured noise. Both processes distort
    the recovered NN distribution and may either dilute or amplify the apparent
    projection bias.
    
    \item Sensitivity limits.  
    Finite sensitivity and noise variations lead to incompleteness, particularly
    for faint or compact cores near the detection threshold. This effectively reduces
    the observable sample size and increases the measured mean separation by removing
    the (smallest) pairs. For a randomly thinned population, the average NN separation
    scales roughly as $\langle\ell_{\rm 2D}\rangle \propto N_{\rm eff}^{-1/2}$, so even
    moderate incompleteness (10–30\%) can inflate apparent separations by tens of percent.
    
    \item Background confusion and structured emission. 
    Inhomogeneous cloud emission or absorption can produce spatially varying
    completeness across the field, selectively obscuring compact or blended sources in
    crowded regions. This effect preferentially removes the shortest projected
    separations (i.e. those most likely to represent true physical neighbours) thereby
    biasing the apparent spacing distribution towards larger values. Such
    environment-dependent incompleteness acts in the same direction as beam blending,
    but with stronger spatial correlation and greater potential to distort clustering
    statistics near bright ridges or filaments.
    
    \item Field of view and spatial coverage.  
    The effective FoV sets how much of a clump or cloud contributes
    to the statistics and therefore directly influences the apparent sample size, $N$.
    Throughout this work we implicitly assume a single clump roughly enclosed by the observations (e.g. one ALMA primary beam).
    In practice, mosaics and primary-beam attenuation can truncate peripheral
    emission or omit outer cores belonging to the same parent structure. This
    censors the longest separations and narrows the observed spacing distribution. 
    This then means that the observation strategy could matter. 
    Wide mosaics such as \textsc{ALMA-IMF} typically
    deliver $N\!\gtrsim\!100$–200 cores per field \citep{Motte2022}, whereas single–pointing studies
    such as \textsc{ASHES} often yield $N\!\sim\!10$–30 per clump \citep{Morii2024}. Even if the core
    surface density is comparable, a larger FoV raises the measured $N$, and if used
    naively in Eq.~\ref{eq:R_N_SDR}, it increases the inferred correction purely
    because more area was included. Likewise, the mosaic FoV also increases the
    spatial dynamic range, \(\mathrm{SDR}=\mathrm{FoV}/\mathrm{FWHM}_{\rm beam}\),
    pushing $\mathcal{C}(N,\mathrm{SDR})$ closer to its asymptote. Thus $N$ (and $\mathrm{SDR}$)
    are observationally dependent quantities unless defined per a fixed aperture.
    
    \item Idealised morphologies.  
    The models explored here are spherically symmetric or statistically isotropic
    fractals designed to isolate geometric effects. Real star-forming regions,
    however, contain filaments, hubs, and hierarchical networks of substructure.
    Orientation-dependent projection of such anisotropic features can alter the
    observed NN statistics, typically changing $C$ by 10–20\% but potentially more
    for strongly elongated geometries (e.g.~end-on filaments or sheets seen nearly
    face-on). These departures highlight that while our calibration captures the
    dominant sampling and crowding effects, detailed modelling may be required for
    strongly anisotropic morphologies.
    
    \item Intrinsic versus observed sample size.
    Finally, our calibration is expressed in terms of the intrinsic number of objects
    $N$ in the 3D distribution, whereas observations only provide the number of
    detected cores, $N_{\rm obs}$. Incomplete detection (due to sensitivity,
    blending, or limited FoV mentioned above) implies $N_{\rm obs} \leq N$, so using $N_{\rm obs}$ in
    Eq.~\ref{eq:R_N_SDR} systematically underestimates $\mathcal{C}(N,\mathrm{SDR})$ and
    hence the inferred 3D separations. In this sense, corrections based on the
    observed $N_{\rm obs}$ should be regarded as conservative lower limits. A more
    rigorous application would estimate an effective intrinsic $N$ — for example via
    completeness corrections or priors on the core surface density within a fixed
    physical aperture — before evaluating $\mathcal{C}(N,\mathrm{SDR})$.
  
\end{itemize}

These caveats point naturally to several avenues for future work:

\begin{enumerate}

    \item Simulations.  
    Extending the NN projection analysis to full magneto–hydrodynamic
    simulations can test whether the empirical calibration remains valid once
    realistic physics (e.g. gravity, turbulence, feedback, and magnetic fields) shape the
    spatial and kinematic distribution of cores (e.g.\citealp{Lebreuilly2025, Tung2025, Nucara2025}). These tests would also quantify how
    fragmentation geometry and crowding evolve dynamically, allowing the derived
    $\mathcal{C}(N,\mathrm{SDR})$ relation to be linked to physical time-scales and evolutionary
    stages.
    
    \item Forward modelling of observed regions.  
    For individual molecular clouds with modellable geometry (e.g.~filamentary or
    hub–filament systems), tailored mock observations can be used to constrain
    the orientation dependence \(C(N,i)\) as a function of inclination and line-of-sight
    structure.

    \item Finite resolution and beam blending.
    A realistic treatment should ``smooth'' the projected (2D) maps with an
    instrumental beam so that cores are no longer point-like, and then recover
    their distribution using standard, observation-driven tools (e.g.\ \textsc{astrodendro}
    dendrograms, \textsc{getsf}, \textsc{CuTEx}, \textsc{clumpfind}). Beam
    convolution and spatial filtering merge nearby emission peaks, suppress the
    shortest separations, and reduce the effective count of recovered objects,
    driving \(\mathcal{C}(N,\mathrm{SDR})\) towards unity. Systematically varying the taper,
    clean beam, and deblending thresholds in such synthetic pipelines would
    calibrate where the transition occurs and refine the empirical parameter \(S_0\)
    that marks the onset of blending in our model.

    \item Field of view, spatial coverage, and completeness.  
    In this work we treat each field as a single clump, but in practice $N$ also depends on FoV so parameterising the correction by the core surface density $\Sigma_{\rm core}$ or a matched physical aperture provides a more consistent comparison across datasets.
    Practical ways forward include (i) reporting both $N$ and $\Sigma_{\rm core}$,
    along with distance and beam FWHM; (ii) applying $\mathcal{C}(N,\mathrm{SDR})$ using an
    $N$ measured within a matched physical or angular aperture when comparing
    regions with differing FoV; and (iii) re-evaluating $\mathcal{C}$ across nested apertures
    to assess sensitivity to coverage.  Synthetic experiments that vary FoV at fixed
    $\Sigma_{\rm core}$ and $\mathrm{SDR}$ can quantify how coverage-induced
    changes in $N$ propagate into $\mathcal{C}(N,\mathrm{SDR})$ and establish robust
    comparison practices between narrow- and wide-field surveys.  
    Finally, comparisons with numerical simulations of evolving clusters—where
    physical core disruption and migration can modify apparent completeness—could
    help disentangle observational incompleteness from genuine dynamical evolution,
    and motivate a future re-parameterisation of the correction in terms of both
    $\Sigma_{\rm core}$ and evolutionary state.

    \item Incorporating kinematic priors.  
    Future extensions could link cores not only in projected space but also in
    velocity space, using the assumption that objects close in $v_{\mathrm{LSR}}$
    are more likely to be physically associated. Such tests would constrain
    additional priors on the spatial distribution and refine the estimation of
    true 3D separations.
    
    \item Testing line-of-sight reconstruction methods.  
    Because the intrinsic 3D positions are known in our simulations, this framework
    provides an ideal benchmark for evaluating MCMC-based reconstruction approaches
    such as those proposed by \citet{Svoboda2019} and \citet{Traficante2023}.
    
    \item Applications beyond dense cores.  
    Since NN statistics are widely used in stellar clustering, galaxy surveys,
    and large-scale structure analyses, repeating this exercise in those contexts
    can test the generality and universality of the derived projection corrections.
    
\end{enumerate}

\section{Conclusions}
\label{sec_conclusions}

We have quantified how projection and finite resolution bias the NN
statistics commonly used to characterise spatial structure in star-forming
regions. Using idealised three-dimensional distributions, we showed that the
mean ratio of intrinsic to projected NN separations,
$\mathcal{C} \equiv \langle \ell_{\rm 3D}\rangle / \langle \ell_{\rm 2D}\rangle$,
systematically exceeds the simple geometric expectation, $4/\pi$, once the sample
contains more than a handful of objects. As the number of points increases,
projection not only foreshortens individual separations, but it also rewires
the NN network, replacing many true 3D links with new shorter 2D neighbours.
Finite angular resolution adds a second, opposing bias: Smoothing and beam
blending merge intrinsically distinct cores into single sources, effectively
reducing the number of independent points and inflating the apparent
separations. Taken together, these effects erase most of the original
three-dimensional connectivity. The observed 2D NN graph preserves only a
coarse shadow of the underlying structure, and there is no simple global
geometry, which is the only factor that can recover the true 3D network from projected
separations.

We attempted to capture the combined behaviour of the competing effects of crowding versus resolution 
through an empirical relation,
\[
\mathcal{C}(N,\mathrm{SDR}) = \mathcal{C}_\infty\,[1-\exp(-\mathrm{SDR}/S_0)]
\left(\frac{N}{100}\right)^{\beta},
\]
with $\mathcal{C}_\infty = 1.94$, $S_0 = 21.8$, and $\beta = 0.173$. This relation
reproduces the ensemble-averaged simulation results to within
$\sim$10\,\%. Individual realisations, however, can deviate by
$\sim$30–40\,\% on average, and in rare cases by factors of a few, owing to
stochastic variations in fractal geometry and sampling. The relation spans the
physically relevant range $\mathcal{C}\simeq1$–$2.3$ from unresolved, beam-dominated
regimes (where blending drives $\mathcal{C}\!\to\!1$ and the NN network is almost
completely destroyed and rewired) to well-resolved, well-sampled regimes in
which projection-induced rewiring pushes $\mathcal{C}$ towards its asymptotic plateau,
$\mathcal{C}_\infty\simeq2$.
For practical use, this calibration was implemented in the publicly available
\texttt{corespacing3d} package, which provides a convenience function to evaluate
$\mathcal{C}(N,\mathrm{SDR})$ for arbitrary parameter choices \citep{Barnes2026}.

This work represents a first step towards a unified treatment of projection bias
in spatial statistics. By construction it is deliberately simplified, omitting
several important factors—such as sensitivity limits, background confusion,
strongly anisotropic morphologies, and incomplete field coverage—that can
further distort observed core distributions. The caveats we highlighted show that
two-dimensional separations should be interpreted with caution, as they reflect a
convolution of intrinsic structure, sampling, and observational bias, and that
apparent agreement with theoretical scales (e.g.\ a Jeans length) does not
guarantee a one-to-one physical correspondence. Our calibration provides a
foundation for addressing these effects systematically. Future efforts should
incorporate synthetic observations of hydrodynamic simulations, explicit
sensitivity and completeness cuts, and orientation-dependent extensions,
ultimately enabling a more physically consistent comparison between observed and
intrinsic fragmentation scales.

\begin{acknowledgements}
We are grateful to the referee, Nestor Sanchez, for their constructive suggestions.
This work grew out of conversations at the Stellar Origins 2025 meeting in Vienna. I thank the organisers for creating such a productive environment and the many colleagues who offered thoughtful feedback during and after the meeting. RJP acknowledges support from the Royal Society in the form of a Dorothy Hodgkin fellowship.

\end{acknowledgements}

\bibliographystyle{aa}
\bibliography{references}

\begin{appendix}

\section{Dependence on radial profile and sample size}
\label{app:profiles}

\begin{figure*}[t]
  \centering
  \includegraphics[width=\linewidth]{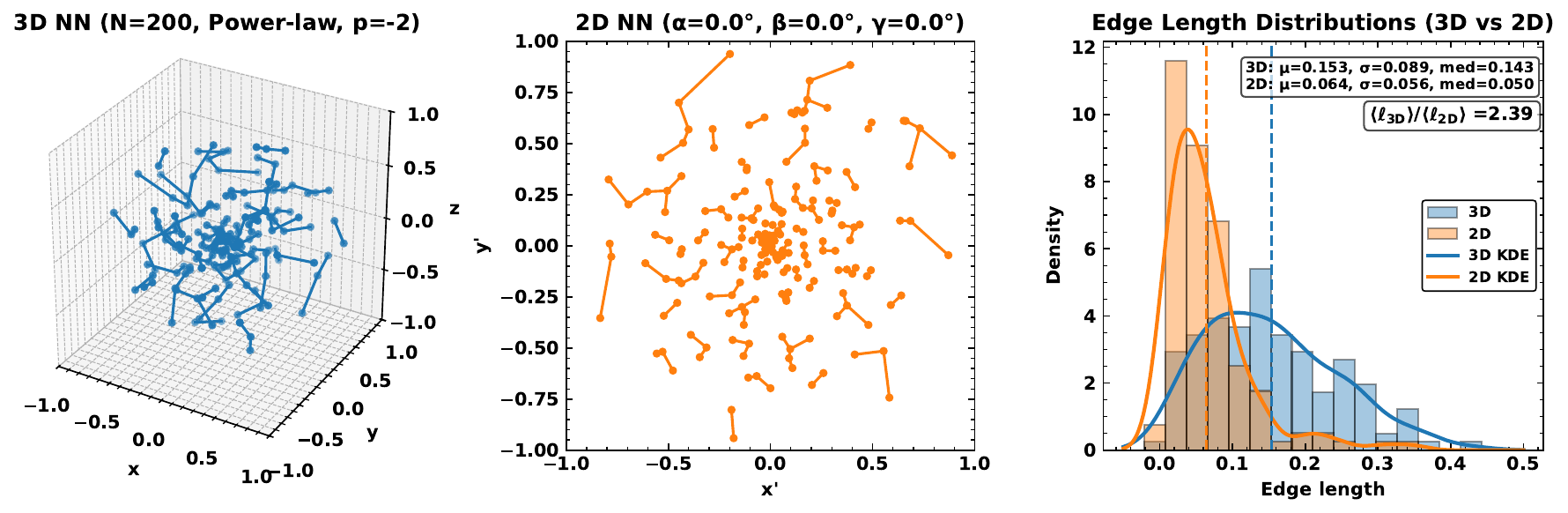} \\
  \includegraphics[width=\linewidth]{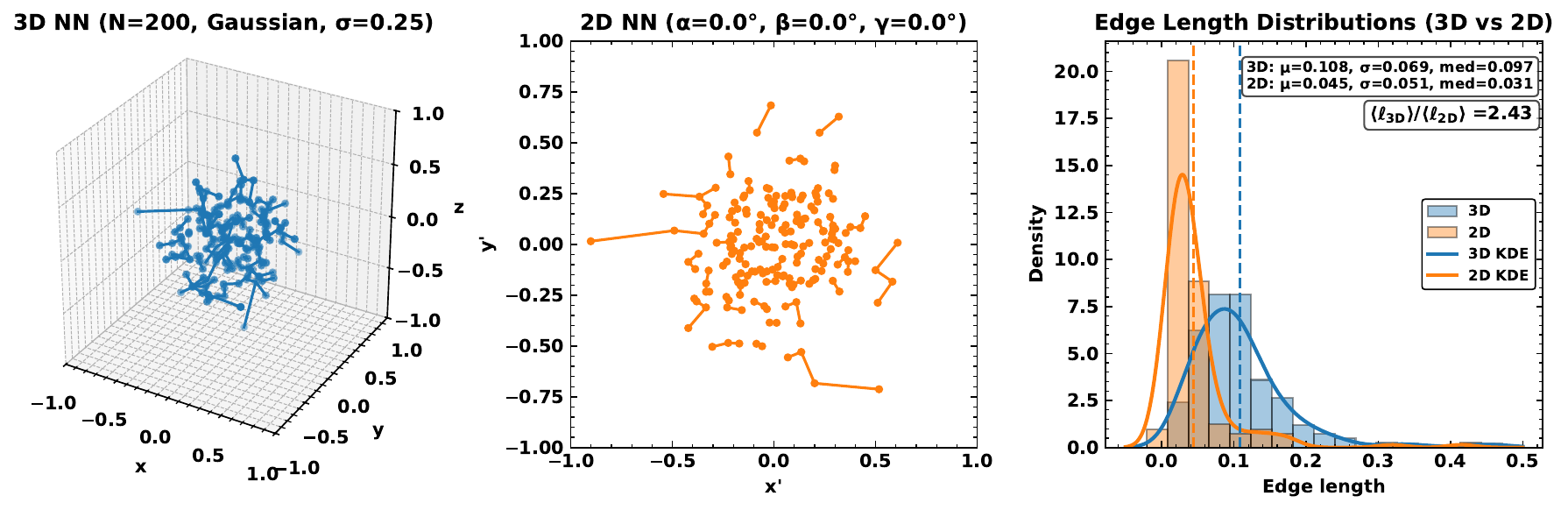} \\
  \includegraphics[width=\linewidth]{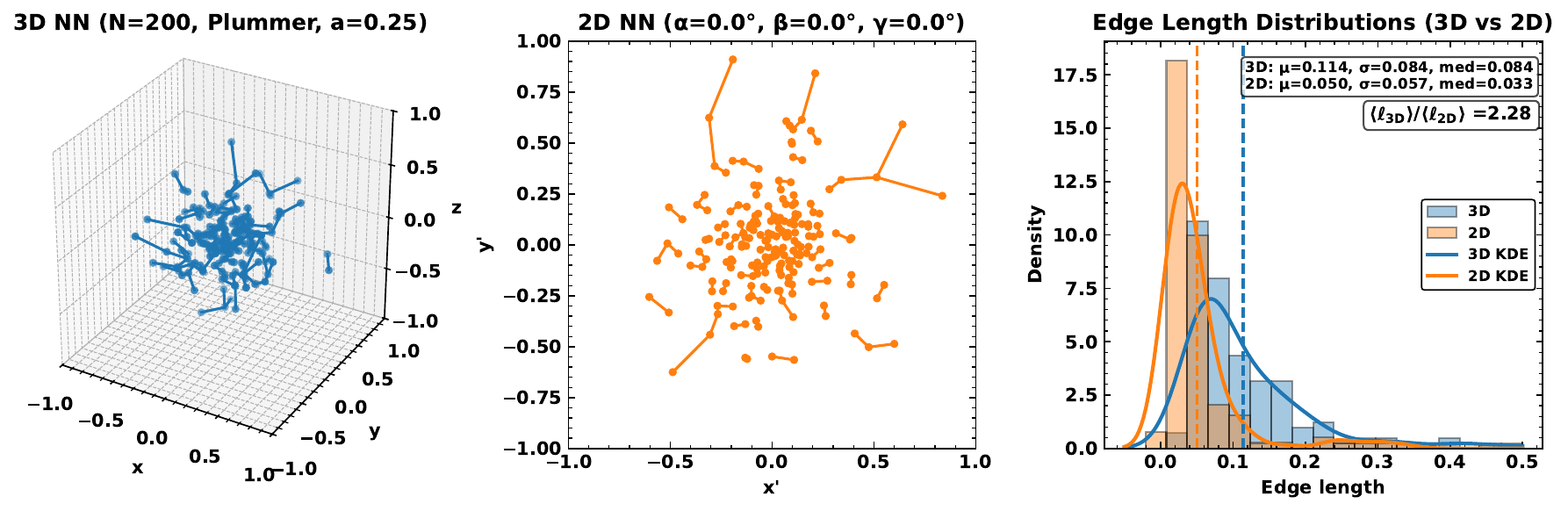} \\
  \caption{
    Nearest–neighbour comparisons for $N=200$ points drawn from three
    centrally concentrated spherical profiles within $R=1.0$. 
    Top: Power law with slope $p=-2$.
    Middle: truncated Gaussian with $\sigma=0.25$. 
    Bottom: Plummer with scale radius $a=0.25$. 
    Each row shows the 3D NN graph (left), the 2D projection (centre), and the
    corresponding distributions of NN edge lengths (right). Projection shortens
    apparent separations and rewires most nodes to new, closer neighbours in 2D,
    with mean ratios $\ell_{\rm 3D}/\ell_{\rm 2D}\!\sim\!2.3$–2.4, well above
    the geometric baseline $4/\pi\simeq1.27$.
  }   
  \label{fig:nn_profiles}
\end{figure*}

While the main text focuses on the uniform-density case as a clean baseline for
understanding projection effects, the underlying behaviour is robust across a
wide range of centrally concentrated spherical profiles (see Fig.\,\ref{fig:nn_profiles}).  
Here we summarise the corresponding results for Gaussian, power-law, and
Plummer models, and verify that the scaling with sample size $N$ and the
characteristic 3D-to-2D ratio remain nearly invariant.  
These tests confirm that the projection bias is governed primarily by crowding
and sample size rather than the specific form of the radial density profile.

Cores are drawn from isotropic distributions with a prescribed radial density
profile within a sphere of radius $R$:
\begin{itemize}
  \item Gaussian: $\rho(r)\propto\exp[-r^2/(2\sigma^2)]$, optionally truncated at $R$, where $\sigma$ sets the Gaussian scale length.
  \item Power law: $\rho(r)\propto r^{p}$ for $0\le r\le R$ with $-3<p<0$ (where $p=0$ is the uniform distribution).
  \item Exponential: $\rho(r)\propto\exp(-r/r_0)$, where $r_0$ sets the exponential scale length.
  \item Plummer: $\rho(r)\propto(1+r^2/a^2)^{-5/2}$, with $a$ the Plummer softening radius.
\end{itemize}
These models are spherically symmetric by construction; orientation has no
effect on their ensemble statistics. In the analysis that follows, we focus on
the uniform case, which provides a transparent baseline for isolating the
effects of projection and finite resolution. The remaining profiles yield
quantitatively similar trends, with modest variations in absolute scaling due to
their differing central concentrations, and are presented for comparison in
Appendix~\ref{app:profiles}.

The mean NN ratios for these models are
\[
\frac{\langle\ell_{\rm 3D}\rangle}{\langle\ell_{\rm 2D}\rangle} \simeq
\begin{cases}
2.43 & \text{(Gaussian)},\\
2.39 & \text{(power law)},\\
2.28 & \text{(Plummer)},
\end{cases}
\]
and they are essentially indistinguishable from the uniform case.  
Median ratios are even larger, confirming that the global correction is not
sensitive to the radial density profile.  
Although central concentration shifts the absolute NN scale to smaller values,
the 3D/2D ratio remains nearly constant because projection is dominated by
neighbour reassignment rather than simple foreshortening.

Only $\sim$20\% of true 3D neighbours are recovered after projection, and
fewer than 10\% of points retain the same neighbour, independent of profile.
This invariance demonstrates that the mean correction factor is driven by the
stochastic geometry of crowding, not by specific density laws.

\section{Nearest--neighbour versus minimum spanning tree analysis}
\label{appendix:discussion_nn_mst}

\begin{figure*}[t]
  \centering
  \includegraphics[width=\linewidth]{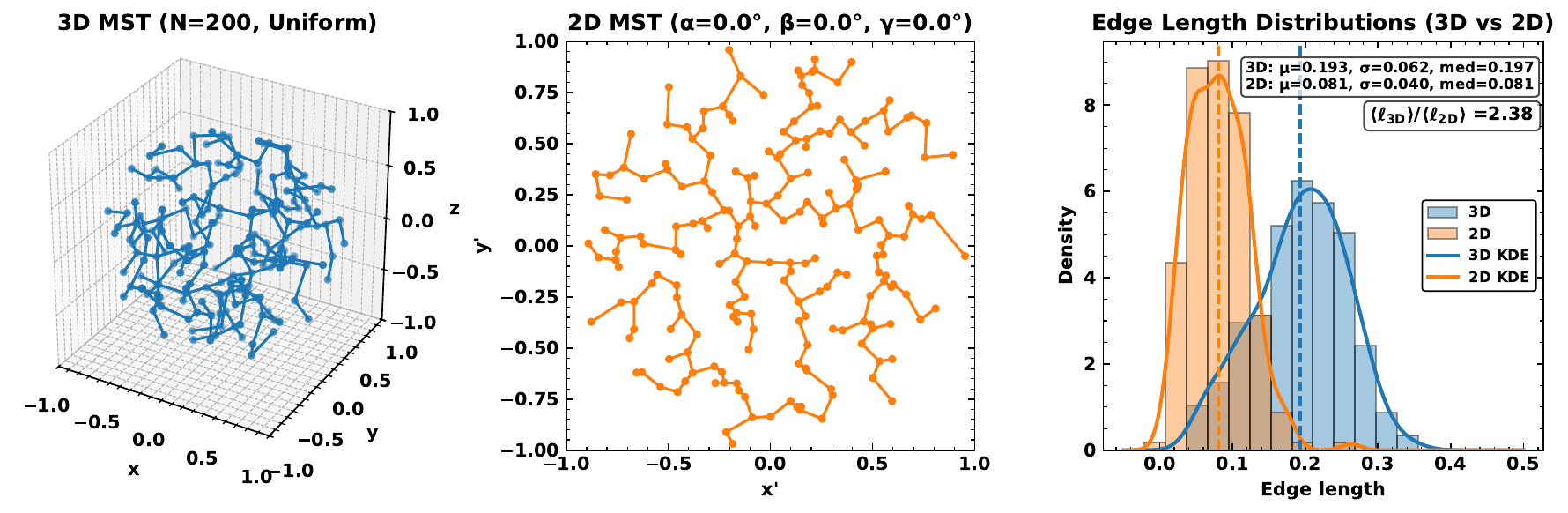}
  \caption{
  Minimum spanning tree comparison for a uniform spherical distribution of
  $N{=}200$ points within radius $R{=}1.0$. 
  Left: MST network constructed in 3D. 
  Centre: MST built from the projected 2D positions (no rotation). 
  Right: Distributions of MST edge lengths in 3D (blue) and 2D (orange),
  with medians (dashed lines) and kernel density estimates (KDEs) overplotted.
  The total MST length decreases from $38.4$ (3D) to $16.1$ (2D),
  corresponding to a mean ratio
  $\langle \ell_{\rm 3D}/\ell_{\rm 2D}\rangle \simeq 2.38$,
  well above the geometric expectation $(4/\pi)\simeq1.27$.
  Only $\sim$25\% of 3D edges are recovered in 2D (Jaccard similarity $0.14$),
  indicating that projection not only foreshortens MST edges but also rewires
  the network topology—though less severely than for the
  NN graph (cf.\ Fig.~\ref{fig:nn_uniform}).
  }
  \label{fig:mst_uniform}
\end{figure*}

It is worth mentioning the relationship between NN and MST analyses since both are
widely used to quantify spatial structure in core and stellar populations.
Although they are often treated interchangeably, they probe different aspects of
clustering and respond differently to projection.

The NN graph captures local proximity, linking each object to its single
closest companion and thus tracing the characteristic fragmentation spacing.
The MST, by contrast, connects all objects into a single loop-free network of
minimum total length, balancing short and long edges to preserve global
connectivity. MST-based measures therefore characterise the overall
geometry and hierarchical organisation of a region rather than its immediate
core separations.

As shown in Fig.\,\ref{fig:mst_uniform}, in a uniform 3D sphere with $N{=}200$ points, the total MST length decreases
from ${\sim}38.4$ in 3D to ${\sim}16.1$ after projection, corresponding to a
mean edge-length ratio
$\langle \ell_{\rm 3D}/\ell_{\rm 2D}\rangle \simeq 2.38$
(see Fig.~\ref{fig:mst_uniform}),
substantially exceeding the geometric expectation of $(4/\pi)\simeq1.27$.
The overlap between the 3D and 2D MST edge sets is modest
($\simeq$25\% of 3D edges recovered; Jaccard similarity $0.14$),
and the median 3D edge length ($\tilde{\ell}_{\rm 3D}=0.197$)
is about 2.4 times larger than the median 2D edge length
($\tilde{\ell}_{\rm 2D}=0.081$).
The NN network exhibits an even stronger contraction, with
$\langle \ell_{\rm 3D}/\ell_{\rm 2D}\rangle\simeq 2.54$
and only $\sim$22\% of edges retained (Jaccard $0.12$),
confirming that both statistics are reshaped by projection.

The relative stability of the MST reflects its global constraint: long
connections remain even after projection, so its mean ratio scales closer to a
geometric expectation, whereas the NN network is dominated by local
reassignment. In this sense, MST statistics are more robust but less sensitive
to the smallest physical separations, while NN statistics are more affected by
projection but directly trace the local fragmentation scale once corrected.

In observational work, MST-based spacing measures are common, as in the ASHES
\citep[e.g.][]{Morii2024} and ALMAGAL \citep{Molinari2025} surveys, where the
mean MST edge length is used to quantify global fragmentation. However, the
MST includes a mixture of short intra-cluster and long inter-cluster links,
which dilutes the local spacing signature that NN analyses capture more cleanly.

For studies aiming to compare measured separations with theoretical scales such
as the Jeans length, the sonic scale, or magnetic critical scales, NN-based
statistics are therefore the most appropriate choice. The empirical correction
derived here applies explicitly to NN separations, but the underlying projection
bias arises from the same 3D geometry. Tests such as those above suggest that
the same sub-linear, $N$-dependent behaviour roughly extends to MSTs as well,
albeit with a smaller amplitude. Thus, while the corrections presented here
should be applied directly to NN-based analyses, they can also provide a useful
first-order approximation for MST-based studies when the underlying spatial
distribution and sampling density are similar.

\section{Dependence of the projection ratio on fractal dimension}
\label{app:ratio_vs_D}

\begin{figure}
\centering
\includegraphics[width=\columnwidth]{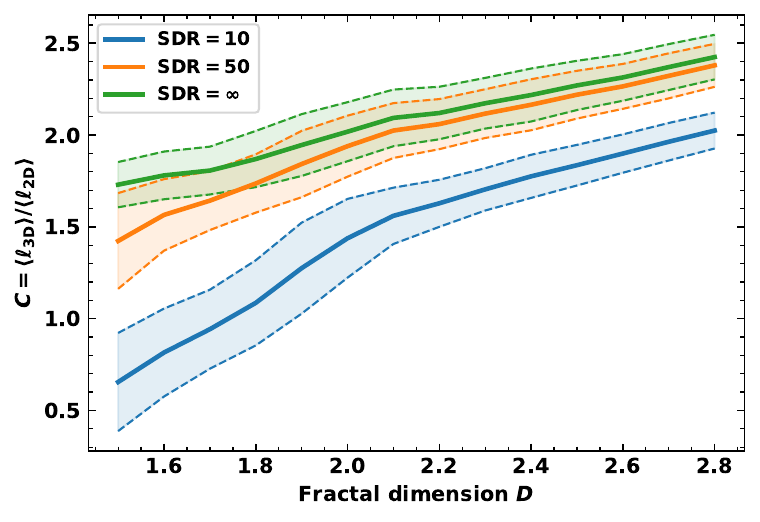}
\caption{
Projection ratio $\mathcal{C} = \langle \ell_{\rm 3D} \rangle / \langle \ell_{\rm 2D} \rangle$
as a function of fractal dimension $D$ for $N=200$, evaluated at three spatial
dynamic ranges: $\mathrm{SDR}=10$ (blue), $\mathrm{SDR}=50$ (orange), and the
effectively infinite--resolution limit $\mathrm{SDR}\rightarrow\infty$ (green).
Lower values of $D$ correspond to more strongly clustered, filamentary
(more ``fractal'') structures, while higher values of $D$ approach smoother,
more space--filling morphologies.
Solid lines show ensemble means, while shaded regions indicate the $1\sigma$
scatter across Monte Carlo realisations.
The projection ratio increases monotonically with $D$ and approaches an
approximately asymptotic value at high $D$, with finite spatial resolution
systematically reducing $\mathcal{C}$ at all fractal dimensions.
}
\label{fig:ratio_vs_D_fractals}
\end{figure}

To further characterise the dependence of the projection ratio
$\mathcal{C}$
on the underlying spatial structure, we examine its variation as a function
of fractal dimension $D$ at fixed sample size and spatial resolution.
Figure~\ref{fig:ratio_vs_D_fractals} shows $\mathcal{C}(D)$ for fractal realisations with
$N=200$, evaluated at three representative spatial dynamic ranges:
$\mathrm{SDR}=10$, $\mathrm{SDR}=50$, and the effectively infinite--resolution
limit $\mathrm{SDR}\rightarrow\infty$.
Each curve represents an ensemble average over Monte Carlo realisations, with
the shaded regions indicating the $1\sigma$ scatter arising from stochastic
sampling and variations between individual fractal realisations.

At fixed $\mathrm{SDR}$, the projection ratio increases monotonically with
fractal dimension.
Low values of $D$, corresponding to highly filamentary and strongly clustered
morphologies, yield systematically smaller values of $\mathcal{C}$, while progressively
more space--filling structures produce larger ratios.
The increase with $D$ becomes weak at higher fractal dimensions, with $\mathcal{C}$
approaching an approximately asymptotic value for $D \gtrsim 2.4$--2.6,
consistent with the behaviour expected for nearly space--filling (smooth)
three--dimensional distributions.

Finite spatial resolution systematically suppresses the projection ratio by
blending small--scale structure.
As a result, lower--$\mathrm{SDR}$ curves are shifted to smaller values of $\mathcal{C}$
at all $D$, and the high--$D$ asymptotic limit is recovered only for
sufficiently large spatial dynamic range.
Over the range $D=1.5$--2.8 explored here, $\mathcal{C}$ spans approximately
0.67--2.03 for $\mathrm{SDR}=10$, 1.43--2.38 for $\mathrm{SDR}=50$, and
1.72--2.43 in the high--resolution limit.

An additional feature is that, at the lowest fractal dimensions and spatial
dynamic ranges, the dependence on $D$ and $\mathrm{SDR}$ is not fully separable.
For highly filamentary, strongly clustered distributions (low $D$), the intrinsic
nearest--neighbour separations are dominated by compact substructures; when
combined with poor spatial resolution (low $\mathrm{SDR}$), beam blending efficiently
merges these compact groupings, effectively erasing the small--scale clustering
that distinguishes low--$D$ morphologies. In this regime, $\mathcal{C}$ is suppressed
more strongly than would be expected from either low $D$ or low $\mathrm{SDR}$ alone,
indicating coupled behaviour between structure and resolution when both are extreme.

Overall, varying $D$ primarily shifts the normalisation of $\mathcal{C}$ and drives a
monotonic increase that saturates for $D\gtrsim2.4$ to 2.6, while the principal trends
with $N$ and $\mathrm{SDR}$ established in the main text remain unchanged. The coupled
low--$D$, low--$\mathrm{SDR}$ regime simply delineates where finite resolution can erase
the structural differences that otherwise distinguish filamentary morphologies.

\end{appendix}

\end{document}